\PassOptionsToPackage{dvipsnames}{xcolor}
\PassOptionsToPackage{oneside}{geometry}
\documentclass[twocolumn, twocolappendix]{aastex701}
\received{\today}
\shorttitle{Charting the Galactic Underworld I}
\shortauthors{Wagg et al.}
\graphicspath{{figures/}}

\usepackage{lipsum}
\usepackage{physics}
\usepackage{multirow, makecell}
\usepackage{xspace}
\usepackage{natbib}
\usepackage{fontawesome5}
\usepackage{wrapfig}
\usepackage[figuresright]{rotating}
\usepackage{subcaption}
\usepackage{booktabs}
\usepackage{siunitx}

\usepackage[hang,flushmargin]{footmisc}

\usepackage{styles/todo_notes}
\usepackage{styles/abbreviations}

\makeatletter
\renewcommand{\unit}[1]{%
    \,\mathrm{#1}\checknextarg}
\newcommand{\checknextarg}{\@ifnextchar\bgroup{\gobblenextarg}{}}
\newcommand{\gobblenextarg}[1]{\,\mathrm{#1}\@ifnextchar\bgroup{\gobblenextarg}{}}
\makeatother

\newif\ifstartedinmathmode
\newcommand{\msun}{%
  \relax\ifmmode\startedinmathmodetrue\else\startedinmathmodefalse\fi
  {\ifstartedinmathmode\unit{M_{\odot}}\else$\unit{M_{\odot}}$\fi}\xspace%
}

\newif\ifstartedinmathmode
\newcommand{\rsun}{%
  \relax\ifmmode\startedinmathmodetrue\else\startedinmathmodefalse\fi
  {\ifstartedinmathmode\unit{R_{\odot}}\else$\unit{R_{\odot}}$\fi}\xspace%
}

\definecolor{codecolour}{HTML}{905cc4}

\defcitealias{underworld}{S22}
\defcitealias{Olejak+2020:2020A&A...638A..94O}{O20}
\defcitealias{Fryer+2012:2012ApJ...749...91F}{F12}
\defcitealias{Hobbs+2005}{Hobbs05}
\defcitealias{Mandel+2020:2020MNRAS.499.3214M}{MM20}
\defcitealias{Maltsev+2025:2025AA...700A..20M}{Maltsev25}

\newcommand{\median}[1]{\text{med}[#1]}

\newenvironment{conclusions}{\begin{enumerate}}{\end{enumerate}}

\newcommand{\conclusion}[2]{\item \textbf{#1}\\#2}
\newcolumntype{T}[1]{>{\centering\arraybackslash}p{#1}}

\newcommand{\eqcol}[1]{\makebox[1cm][c]{#1}}   

\begin{document}


\title{Charting the Galactic Underworld I:\\Comprehensive simulations of the kinematics, rates, and demographics of Milky Way black holes}

\newcommand{\cca}{\affiliation{Center for Computational Astrophysics, Flatiron Institute, 162 Fifth Ave, New York, NY, 10010, USA}}
\newcommand{\cmu}{\affiliation{McWilliams Center for Cosmology and Astrophysics, Department of Physics, Carnegie Mellon University, Pittsburgh, PA 15213, USA}}
\newcommand{\arizona}{\affiliation{University of Arizona, Department of Astronomy \& Steward Observatory, 933 N. Cherry Ave., Tucson, AZ 85721, USA}}
\newcommand{\UW}{\affiliation{Department of Astronomy, University of Washington, Seattle, WA, 98195}}
\newcommand{\UCB}{\affiliation{Department of Astronomy, University of California, Berkeley, Berkeley, CA 94720, USA}}

\author[0000-0001-6147-5761]{Tom Wagg}
\cca{}
\email{tomjwagg@gmail.com}

\author[0000-0001-5228-6598]{Katelyn Breivik}
\cmu{}
\email{kbreivik@andrew.cmu.edu}

\author[0000-0003-0872-7098]{Adrian~M.~Price-Whelan}
\cca{}
\email{aprice-whelan@flatironinstitute.org}

\author[0000-0002-6718-9472]{Mathieu Renzo}
\arizona{}
\email{mrenzo@arizona.edu}

\author[0000-0002-1264-2006]{Julianne J.\ Dalcanton}
\cca{}
\UW{}
\email{jdalcanton@flatironinstitute.org}

\author[0000-0002-0287-3783]{Natasha S.\ Abrams}
\UCB{}
\email{nsabrams@berkeley.edu}

\correspondingauthor{Tom Wagg}
\email{twagg@flatironinstitute.org, tomjwagg@gmail.com}

\begin{abstract}
    With upcoming data from \textit{Roman}, \gaia DR4, and spectroscopic surveys, we will soon have an unprecedented dataset of Milky Way black holes (BHs) to constrain their formation and evolution.
    To prepare, we simulate the intrinsic population of Milky Way BHs with \cogsworth, self-consistently accounting for their binary evolution and trajectories through the Galactic potential. We report the rate, demographics, and kinematics of these BHs, and their sensitivity to 32 variations in binary evolution, supernova physics, and Galactic potentials.
    In the fiducial model, ${\sim}1.7 \times 10^8$ BHs have formed in the Milky Way (though this total spans an order of magnitude across our variations), where the vast majority (${\sim}91\%$) are currently isolated and ${\sim}3\%$ have escaped the Galaxy.
    Most of the ${\sim}10^7$ BHs in binaries have another BH or a white dwarf companion, but ${\sim}10^5$ retain a luminous stellar companion.
    BHs are distributed more diffusely than visible stars, with a scale height around ${\sim}2.5\times$ larger. BH masses correlate with present-day location: the most massive BHs are preferentially close to the Galactic plane. This correlation is especially strong for BH–star binaries, which separate into tight, low-mass post-common-envelope systems and wide, high-mass non-interacting ones. The BH mass distribution and kinematics are highly sensitive to the remnant mass prescription and natal kick model, so observations could constrain explodability criteria and BH kicks. Accounting for the time-evolution of the Galactic potential more than doubles the escape fraction and increases the bound population's scale height by ${\sim}20\%$, whilst neglecting binary interactions overestimates it by 30\%.
\end{abstract}


\section{Introduction}

Stellar-mass black holes (BHs) are the endpoints of some massive stars, formed either through a core-collapse supernova (SN), or via direct collapse \citep[e.g.,][]{Ruffini+1971:1971PhT....24a..30R, Burrows+2025:2025ApJ...987..164B}. The rates and demographics of these BHs are imprinted with information about their progenitors' evolution and thus offer insights about (binary) stellar evolution. Populations of BHs can also help reveal aspects of SN physics, such as the degree of asymmetry in SN explosions, or the relation between pre-SN core masses and final BH masses \citep[e.g.,][]{Timmes+1996}.

To date, the vast majority of BHs have been detected as part of a binary system. The first BHs were discovered in X-ray binaries \citep[XRBs, ][]{Webster+1972:1972Natur.235...37W, Bolton+1972:1972Natur.235..271B} as a result of accretion from a stellar companion. This method has been used to identify more than 70 BH candidates in the Milky Way \citep{Corral-Santana+2016}\footnote{\url{https://www.astro.puc.cl/BlackCAT}}. With the advent of gravitational-wave (GW) observations, the number of observed BHs in binaries has rapidly grown. As of GWTC-5, the LIGO-Virgo-KAGRA collaboration has reported the detection of 391 GW mergers, corresponding to ${\sim}$535 measurements of an individual component mass confidently above $5 \msun$ \citep{Abbott+2019_GWTC1, Abbott+2020_GWTC2, GWTC_2_1, Abbott+2023:2023PhRvX..13d1039A_GWTC3, Abac+2025:2025ApJ...995L..18A_GWTC4,
TheLIGOScientificCollaboration+2025:2025arXiv250818079T_GWTC_Open_data, GWOSC_snapshot_v23, TheLIGOScientificCollaboration+2026:2026arXiv260527226T}. Every BH detected via GWs so far is extragalactic; however, future millihertz detectors such as \lisa are expected to detect around a hundred BHs in the Milky Way \citep[e.g.,][]{Nelemans+2001, Wagg+2022}. Three BHs in wide orbits have been discovered in \gaia observations and confirmed spectroscopically \citep{El-Badry+2023:2023MNRAS.518.1057E_GaiaBH1, El-Badry+2023:2023MNRAS.521.4323E_GaiaBH2, Chakrabarti+2023:2023AJ....166....6C, GaiaCollaboration+2024:2024AA...686L...2G_GaiaBH3}, with dozens more expected in \gaia DR4 later this year \citep[e.g.,][]{El-Badry+2023:2023MNRAS.518.1057E_GaiaBH1}. Other candidate BHs have been found from radial velocity measurements of stellar companions in globular clusters \citep[e.g.,][]{Giesers+2018:2018MNRAS.475L..15G,Giesers+2019:2019A&A...632A...3G, Whitaker+2026:2026arXiv260618350W} or the LMC \citep{Shenar+2022:2022NatAs...6.1085S}.

Despite most BHs being detected in binaries, the majority of the intrinsic population are predicted to be isolated \citep[e.g.,][hereafter \citetalias{Olejak+2020:2020A&A...638A..94O}]{Olejak+2020:2020A&A...638A..94O}. Most BH progenitors are \textit{formed} in binaries \citep[e.g.,][]{Offner+2023:2023ASPC..534..275O}, but the majority of these binaries are disrupted by the first supernova \citep[e.g.,][]{Renzo+2019:2019A&A...624A..66R, Wagg+2025:2025OJAp....8E..85W, Wagg+2025:2025AJ....170..192W} and many others merge as a result of unstable mass transfer \citep[e.g.,][]{Hjellming+1987:1987ApJ...318..794H, Webbink+1984:1984ApJ...277..355W}. Indeed, below we find that even in the absence of asymmetric supernova natal kicks, around 68\% of BHs in the Milky Way would be isolated.

The population of isolated BHs is more difficult to detect due to the lack of strong electromagnetic counterparts or gravitational-wave signatures. The most promising method for detecting isolated BHs is through gravitational microlensing \citep{Paczynski+1986:1986ApJ...304....1P}. When a BH passes in front of a background star, the warping of space--time ``lenses'' the star, leading to both an astrometric shift and photometric amplification. From this lensing, one can derive the mass, distance, and proper motion of the lens \citep[e.g.][]{Dominik+2000:2000ApJ...534..213D}. So far, only 1 isolated BH has been detected via microlensing \citep{Lam+2022:2022ApJ...933L..23L, Sahu+2022:2022ApJ...933...83S, Mroz+2022:2022ApJ...937L..24M, Lam+2023:2023ApJ...955..116L, Sahu+2025:2025ApJ...983..104S}, but 100s more are expected with observations from the \romanFull \citep[e.g.,][]{Lam+2022:2022ApJ...933L..23L}. Furthermore, the accretion of dense material from the interstellar medium may produce detectable X-ray emission from nearby BHs \citep[e.g.,][]{Shvartsman+1971:1971SvA....15..377S,Gaggero+2017:2017PhRvL.118x1101G,Tsuna+2018:2018MNRAS.477..791T}.

Isolated BHs offer complementary constraints on massive stellar evolution compared to the bound population. The formation of a BH via a core-collapse SN (CCSN) is expected to impart a velocity kick on the system. This kick is a result of a combination of symmetric mass loss \citep{Blaauw+1961, Boersma+1961} and asymmetry in the SN explosion, both in baryons \citep[e.g.,][]{Shklovskii+1970:1970SvA....13..562S,Lyne+1994,Janka+2013:2013MNRAS.434.1355J,Janka+2017:2017ApJ...837...84J, Burrows+2024:2024ApJ...963...63B} and neutrino emission \citep[e.g.,][]{Janka+1994:1994A&A...290..496J,Burrows+1996:1996PhRvL..76..352B}. In the majority of cases, these kicks are strong enough to disrupt a binary orbit \citep[e.g.,][]{DeDonder+1997:1997A&A...318..812D, Renzo+2019:2019A&A...624A..66R}. Therefore, the requirement of the BH remaining in a bound system applies a strong selection effect, favouring tight pre-SN systems and those with weak SN kicks. BHs may also form via direct-collapse, or fail to eject their envelopes in CCSNe, causing most of the stellar material to fall back onto the collapsing core \citep[e.g.,][]{Kochanek+2008:2008ApJ...684.1336K, De+2026:2026Sci...391..689D}. In these cases any natal kick from baryons is negated and thus the velocity imparted to these BHs is weak.

The spatial distribution of BHs is significantly different from the visible stellar population in the Milky Way as a result of SN kicks \citep[e.g.,][hereafter \citetalias{underworld}]{underworld}. The present-day location and velocity of BHs therefore holds imprints of their prior evolution \citep[e.g.,][]{Andrews+2022:2022ApJ...930..159A, Vigna-Gomez+2023:2023ApJ...946L...2V}. Observations of isolated BHs and their kinematics can therefore inform our understanding of SN explosions and massive stellar evolution.

The Milky Way population of BHs has been explored in several previous works \citep[e.g.,][]{Shapiro+1983:1983bhwd.book.....S,vandenHeuvel+1992:1992eocm.rept...29V, Brown+1994:1994ApJ...423..659B}, but with no self-consistent treatment of binary interactions and galactic orbit integration. In recent years, \citet{Wiktorowicz+2019:2019ApJ...885....1W} simulated binary populations with \texttt{StarTrack} \citep{Belczynski+2008}, exploring how the population of stellar-mass BHs varies for different assumptions regarding initial distributions and natal kicks. Their work assumes a single metallicity population and focuses on a more general population rather than specifically on the Milky Way. \citetalias{Olejak+2020:2020A&A...638A..94O} also uses \texttt{StarTrack} simulations, producing a full synthetic catalogue of Milky Way BHs. They perform one variation of common-envelope physics assumptions, but do not integrate the positions of BHs through the galactic gravitational potential. 

Most recently, \citet{underworld}, hereafter \citetalias{underworld}, simulated the ``Galactic Underworld'', making predictions for the spatial distribution of Milky Way BHs. Their work fully integrates the orbits of BHs through a galactic potential until present day, which is important given that BHs receive natal kicks and are long-lived. However, their work assumes that all BHs are formed from single stars and have a mass of $7.8\msun$, and does not quantify the impact of uncertainties in stellar evolution.

In this paper, we build upon previous works by leveraging the open-source code \cogsworth \citep{Wagg+2025:2025JOSS...10.7400W, cogsworth_apjs}, which enables self-consistent binary population synthesis and galactic dynamics simulations. With \cogsworth, we are able to consistently evolve a full population of BHs, accounting for their binary interactions, formation in a detailed Milky Way star formation history, and trajectories through a full Galactic gravitational potential. We report the rate, demographics, and kinematics of the Milky Way population of BHs, in addition to correlations between their mass and location. We explore the sensitivity of our results to variations in our assumptions regarding (binary) stellar evolution and the Galactic potential.

This paper is structured as follows: in Section~\ref{sec:methods} we outline the simulation framework and fiducial assumptions. We consider the fiducial population of Milky Way BHs in Section~\ref{sec:fid_results}, before exploring the impact of the various model variations in Section~\ref{sec:variations}. We examine the population of BHs that remain in bound systems with a companion at present day in Section~\ref{sec:bh_in_binaries}. Finally, we discuss the implications of results for the observational landscape of BHs in Section~\ref{sec:discussion}, before concluding in Section~\ref{sec:conclusions}. The code for reproducing the simulations and figures for this work is available \href{https://github.com/TomWagg/underworld/}{on GitHub}\footnote{\url{https://github.com/TomWagg/underworld/}} and \href{https://doi.org/10.5281/zenodo.21536616}{Zenodo}\footnote{\url{https://doi.org/10.5281/zenodo.21536616}}.

\section{Methods}\label{sec:methods}

We use \cogsworth \citep[][v4.0.2]{Wagg+2025:2025JOSS...10.7400W, cogsworth_apjs} to simulate the Galactic population of BHs. \cogsworth uses \texttt{COSMIC} \citep[][v4.1.1]{COSMIC} to rapidly synthesize populations of binary stars and \gala \citep[][v1.11.1]{Gala} to integrate the subsequent orbits of the (potentially unbound) binaries in model galactic potentials. A detailed description of \cogsworth can be found in \citet{cogsworth_apjs}, but we give a brief overview of our approach here. The subsequent subsections outline our fiducial assumptions regarding initial stellar population sampling (\S~\ref{sec:initial_sampling}), the star formation history of the Galaxy (\S~\ref{sec:sfh}), stellar and binary physics (\S~\ref{sec:fiducial-stellar}), and the Galactic potential (\S~\ref{sec:fiducial-pot}). We summarise these settings in Table~\ref{tab:fiducial_settings}.

\subsection{Overview}

We simulate $6 \times 10^8 \msun$ of stellar mass, representing approximately 1/100th of the total stellar mass of the Milky Way \citep{Licquia+2015}. We sample the initial binary stellar population from this stellar mass, discarding any binaries in which the initially more massive star is less than $4 \msun$. We are interested in studying BHs and therefore binaries with a total mass below $8 \msun$ are not relevant to this study. This limit is quite conservative for BHs, but ensures that the sample contains all binaries that can reach core-collapse, and thus means our dataset is also relevant for future work concerning neutron stars (NSs).

We sample Galactic birth times, locations, velocities, and metallicities for each of these binaries using a distribution-function-based star formation history model based on \citep{Sanders+2015:2015MNRAS.449.3479S}. We evolve this initial population from birth until present day with \cogsworth, retain any system in which a BH is formed, and track its full evolutionary history and trajectory through the Galaxy. In total, we simulate the evolution of around 53 million BHs for our 32 parameter variations.

\subsection{Initial stellar population sampling}\label{sec:initial_sampling}

\begin{table*}
    \centering
    \begin{tabular}{lll}
\toprule
Setting name & Fiducial choice & Reference \\
\midrule
\multicolumn{3}{l}{\textbf{Supernova natal kicks}} \\
\quad Core-collapse SN kicks & lognormal ($\mu = 5.6, \sigma = 0.68$) & \citet{Disberg+2025:2025ApJ...989L...8D} \\
\quad Electron-capture SN kicks & Maxwellian $(\sigma=20\unit{km}{s^{-1}})$ & \citet{Igoshev+2020} \\
\quad BH kicks & As above, rescaled by fallback fraction & \citet{Fryer+2012:2012ApJ...749...91F} \\
\addlinespace[0.5em]
\hline
\addlinespace[0.5em]
\textbf{Remnant mass prescription} & Delayed explosions after core bounce & \citet{Fryer+2012:2012ApJ...749...91F} \\
\addlinespace[0.5em]
\hline
\addlinespace[0.5em]
\multicolumn{3}{l}{\textbf{Mass transfer}} \\
\quad Mass transfer stability & Determined with critical mass ratio $q_{\rm crit}$ & \citet{Neijssel+2019}\\
\addlinespace[0.5em]
\quad \shortstack[l]{Stable mass transfer efficiency ($\beta$)\\\quad\\\quad} & \shortstack[l]{Fraction of donated mass accepted\\by accretor. Limited to 10x accretor\\thermal timescale, unlimited for giants} & \shortstack[l]{\citet{Kippenhahn+1967:1967ZA.....65..251K} \\ \citet{Schneider+2015:2015ApJ...805...20S}\\\quad} \\
\addlinespace[0.5em]
\quad Angular momentum loss & Isotropic re-emission & \citet{Massevitch+1975:1975MmSAI..46..217M} \\
\addlinespace[0.5em]
\quad \shortstack[l]{Common-envelope efficiency ($\alpha_{\rm CE}$) \\ \quad} & \shortstack[l]{Fraction of orbital energy used to\\unbind envelope, $\alpha_{\rm CE} = 1.0$} & \shortstack[l]{\citet{Webbink+1984:1984ApJ...277..355W}\\ \citet{deKool+1990:1990ApJ...358..189D}} \\
\quad Common-envelope binding energy ($\lambda_{\rm CE}$) & Fits to detailed models from \texttt{STARS} & \citet{Claeys+2014:2014AA...563A..83C} \\
\addlinespace[0.5em]
\hline
\addlinespace[0.5em]
\multicolumn{3}{l}{\textbf{Initial distributions}} \\
\quad Initial mass function & \shortstack[l]{$p(m_1) \propto m^{\alpha_{\rm IMF}} \in [4, 150]$\\with $\alpha_{\rm IMF} = -2.3$} & \citet{Kroupa+2001:2001MNRAS.322..231K} \\
\shortstack[l]{\quad Initial mass ratio distribution \\ \quad} & \shortstack[l]{$f(q) \propto q^{\kappa} \in [0, 1]$ with $\kappa = 0$ \\ \quad} & \shortstack[l]{\citet{Mazeh+1992:1992ApJ...401..265M} \\ \citet{Goldberg+1994:1994AA...282..801G}} \\
\quad Minimum secondary mass & $0.08\unit{M_{\odot}}$ & - \\
\shortstack[l]{\quad Initial period distribution \\ \quad} & \shortstack[l]{$\log_{10}(P_{\rm orb} / \unit{days})^{\pi} \in [0.15, 5.5]$ \\ with $\pi = -0.55$} & \shortstack[l]{\citet{Sana+2012:2012Sci...337..444S} \\ \quad} \\
\quad Initial eccentricity distribution & $f(e) \propto e^{\eta} \in [0, 0.9]$ with $\eta = -0.45$ & \citet{Sana+2012:2012Sci...337..444S} \\
\addlinespace[0.5em]
\hline
\addlinespace[0.5em]
\multicolumn{3}{l}{\textbf{Galactic settings}} \\
\quad \shortstack[l]{Star formation history\\\quad} & \shortstack[l]{Action-based, metallicity-dependent\\for thin and thick disc} & \shortstack[l]{\citet{Sanders+2015:2015MNRAS.449.3479S}\\\quad} \\
\quad Galactic gravitational potential & \texttt{MilkyWayPotential(v2)} of \texttt{Gala} & \citet{Gala} \\
\bottomrule
\end{tabular}
    \caption{Simulation settings for the fiducial model. We describe these settings in detail in Section~\ref{sec:methods} and explore variations of these choices in Section~\ref{sec:variations}.}
    \label{tab:fiducial_settings}
\end{table*}

We draw masses of the primary (initially more massive) star in each binary, $m_1$, following the broken power law initial mass function (IMF) from \citet{Kroupa+2001:2001MNRAS.322..231K}, such that $p(m_1) \propto m_1^{\alpha_{\rm IMF}}$ and sampling only stars with $m_1 > 4 \msun$. The high mass slope (for stars with $m_1 > 1 \msun$) of this IMF has $\alpha_{\rm IMF} = -2.3$. We sample mass ratios, $q \equiv m_2 / m_1$, uniformly in $[q_{\rm min}, 1]$, where $q_{\rm min}$ is set such that the minimum secondary mass is $0.08 \msun$ \citep{Mazeh+1992:1992ApJ...401..265M, Goldberg+1994:1994AA...282..801G}. Eccentricities, $e$, are sampled following \citet{Sana+2012:2012Sci...337..444S} such that $p(e) \propto e^{-0.45}$ with $e \in [0, 0.9]$, where the upper limit is chosen to avoid Roche lobe overflow at pericenter (at birth). For orbital periods, $P$, we follow the distribution of \citet{deMink+2015:2015ApJ...814...58D}, which is an extrapolation of \citet{Sana+2012:2012Sci...337..444S} to lower masses, such that $p(P) \propto (\log_{10}(P / \unit{days}))^{-0.55}$ with $\log_{10}(P / \unit{days}) \in [0.15, 5.5]$. We use a fixed binary fraction of 100\% since we simulate only massive stars which are almost all formed in binaries \citep[e.g.,][]{Moe+2017,Offner+2023:2023ASPC..534..275O}. Additionally, though we do not include truly single stars, our upper orbital period limit allows for a fraction of stars to be effectively single, such that they have no interactions with companions and $\leq 5\%$ mass accretion via stellar winds from a companion. The fraction of effectively single stars in the fiducial population (85\%) is consistent with the findings of \citet{Offner+2023:2023ASPC..534..275O}, for our minimum primary mass of 4 \msun.

\subsection{Star formation history}\label{sec:sfh}

For the Milky Way star formation history, we assume the model of \citet{Sanders+2015:2015MNRAS.449.3479S}. This model provides quasi-isothermal disc distribution functions for stars in the Milky Way as a function of birth time. From this model, we are able to sample birth times, locations, velocities, and metallicities.

The global star formation rate of the Milky Way is given by Eq.\ 10 of \citet{Sanders+2015:2015MNRAS.449.3479S}, with a peak around $10\unit{Gyr}$ ago and an exponential falloff towards present day. Stars that are formed more than 10 Gyr ago are assigned to the thick disc, while younger stars are assigned to the thin disc.

We divide these birth times into 10 bins and assume a quasi-isothermal disc distribution function with the properties of the average birth time in the bin. Specifically, we set $\sigma_{r0}$ and $\sigma_{z0}$, which are the radial and vertical velocity dispersions near the Sun, following Eq.\ 11 of \citet{Sanders+2015:2015MNRAS.449.3479S}. We use the metallicity-age-radius relation presented by \citet{Frankel+2018} (based on APOGEE data) to assign metallicities to each binary based on their initial position and birth time. This model accounts for the inside-out growth of the Galaxy and radial migration of stars. With these inputs, we use \texttt{agama} to sample positions and velocities of binaries \citep{Vasiliev+2019:2019MNRAS.482.1525V}.


A key advantage of this method is that the sampled distribution of positions and velocities is in a steady-state. Integrating the galactic orbits of binaries with these initial conditions forwards in time retains the initial distributions in the absence of perturbations to the orbits. Therefore, any changes in the distribution of positions of BHs in this model are fully a result of supernova kicks.

\subsection{Stellar and binary physics}\label{sec:fiducial-stellar}

\cogsworth uses \texttt{COSMIC} for binary population synthesis. \texttt{COSMIC} is based on the \texttt{BSE} code \citep{Hurley+2000:2000MNRAS.315..543H, Hurley+2002}, which applies fitting formulae from \citet{Tout+1997:1997MNRAS.291..732T} to the single star models of \citet{pols:98}, but with extensive modifications and updated prescriptions based on more recent work \citep[see Section 3 of][]{COSMIC}. The simulations make use of \texttt{COSMIC} v4.1.1 and use the default settings for that version, which we summarise in Table~\ref{tab:fiducial_settings}. In particular, the efficiency of stable mass transfer is defined as, $\beta \equiv \Delta M_{\rm acc} / \Delta M_{\rm don}$, where $\Delta M_{\rm acc}$ is the mass gained by the accretor, and $\Delta M_{\rm don}$ is the mass transferred from the donor. We set $\beta$ such that the amount of mass accreted during Roche-lobe overflow is limited to 10x the thermal rate of the accretor for main sequence, Hertzsprung gap and core helium burning stars and unlimited for giant branch stars \citep[e.g.,][]{Kippenhahn+1967:1967ZA.....65..251K, Schneider+2015:2015ApJ...805...20S}. We assume that mass lost during Roche-lobe overflow is lost from the system via isotropic re-emission (as if it is a wind from the secondary) \citep[e.g.,][]{Massevitch+1975:1975MmSAI..46..217M}. We assume that the masses of compact remnants follow the \textit{delayed} prescription provided by \citet{Fryer+2012:2012ApJ...749...91F}.

We determine the stability of mass transfer using the critical mass ratios defined in \citet{Neijssel+2019}. For unstable mass transfer we assume common-envelope events follow the $\alpha$-$\lambda$ prescription \citep{Webbink+1984:1984ApJ...277..355W, deKool+1990:1990ApJ...358..189D} and by default take $\alpha_{\rm CE} = 1$ and use the $\lambda$ prescription from Appendix A of \citet{Claeys+2014:2014AA...563A..83C}, which uses fits of detailed models from the \texttt{STARS} code \citep{Eggleton+1971:1971MNRAS.151..351E,Pols+1995:1995MNRAS.274..964P}.

Asymmetries in SN explosions impart a natal kick on the compact object that is formed, which may unbind a binary orbit, thereby ejecting a secondary star. We assume SN natal kicks for CCSNe are distributed isotropically as a log-normal distribution, with a mean of $\mu = 5.6$ and standard deviation of $\sigma = 0.68$ \citep{Disberg+2025:2025ApJ...989L...8D}.
The magnitude of SN natal kicks for BHs is modulated based on the \citet{Fryer+2012:2012ApJ...749...91F} fallback mass prescription (see discussion of Figure~\ref{fig:kicks-vs-mass}). We assume that stars with a carbon-oxygen core mass that exceeds 38\msun undergo pulsational pair-instability supernovae (PPISN), following \citet{Renzo+2022:2022RNAAS...6...25R} and \citet{Hendriks+2023:2023MNRAS.526.4130H}, and adopt their criteria for pair-instability supernovae (PISN).

\subsection{Galactic potential}\label{sec:fiducial-pot}
By default, we assume the Galactic gravitational potential is static and follows version 2 of the \gala model for the Milky Way potential \citep{Gala}. This model is fit to observations of the Milky Way rotation curve, the shape of the phase-space spiral in the solar neighbourhood and a compilation of recent mass measurements of the Milky Way \citep{Eilers+2019:2019ApJ...871..120E, Darragh-Ford+2023:2023ApJ...955...74D}.

\section{The population of Milky Way BHs}\label{sec:fid_results}

In this Section we examine the fiducial simulation of the population of Milky Way BHs. We calculate the occurrence rate and fraction of BHs in binary systems in Section~\ref{sec:fid_rates_binarity}, before highlighting their spatial distribution and larger scale height in Section~\ref{sec:fid_spatial}. We show that BH mass and position are correlated in Section~\ref{sec:correlations}, then consider how natal kicks have altered BH velocities and created free-floating BHs in Sections~\ref{sec:fid_vels}~\&~\ref{sec:fid_free_floating}.

\subsection{Rates and binarity}\label{sec:fid_rates_binarity}

We define the occurrence rate of BHs as the number of BHs formed per total Milky Way star-forming mass \citep[$\num{6e10}\msun$,][]{Licquia+2015}. We predict an occurrence rate of $1.7 \times 10^{8}$ BHs over the course of the Milky Way's history, where the total mass of the BHs is $1.5\times10^{9}\msun$. These BHs are formed from $\num{6.7e8}$ core-collapse events, where the remaining fraction form NSs, such that the formation rate of NSs is $2.9\times$ higher than BHs.

\begin{figure}
    \centering
    \includegraphics[width=\columnwidth]{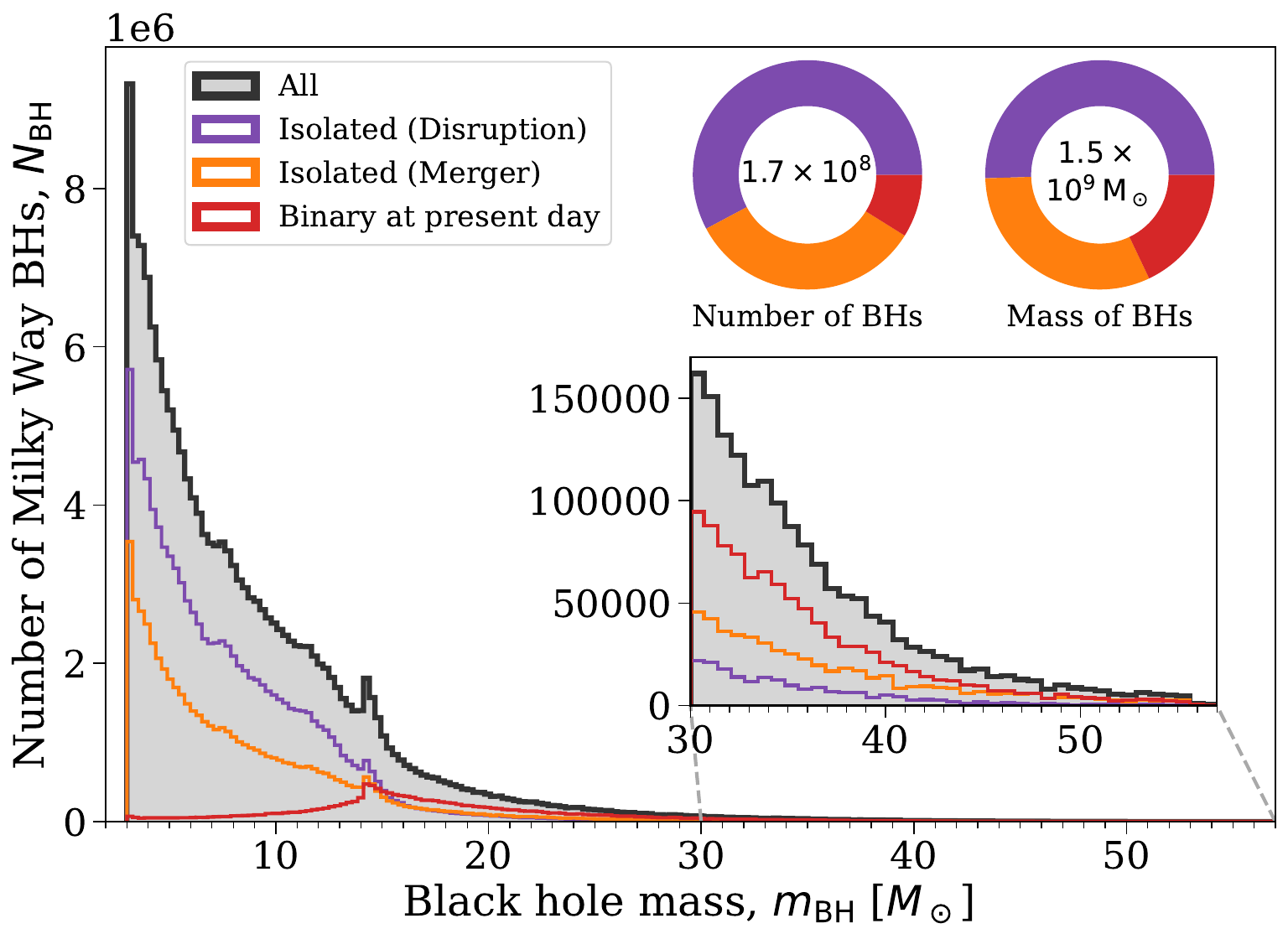}
    \caption{The mass distribution of Milky Way BHs in the fiducial model is shown in black (c.f., Figure~\ref{fig:bh_mass_rmp}a). Coloured histograms separate this distribution based on whether the BH is isolated due to a binary disruption (purple), isolated after a stellar merger (orange), or in a bound binary at present day (red). The inset panel focuses on the high-mass tail, which is dominated by BHs in binaries and those formed from merger products. The two pie charts show the relative contribution of each type to the number and mass budget of BHs respectively. These pie charts are annotated in their centres with the total values for the Milky Way.}
    \label{fig:masses-by-type}
\end{figure}

The distribution of BH masses is concentrated at low masses, peaking at our lowest allowed BH mass of $3 \msun$\footnote{The \citet{Fryer+2012:2012ApJ...749...91F} remnant mass prescription used in the fiducial model only differentiates between BHs and NSs based on this mass cut.}, but has an extended tail. The black histogram in Figure~\ref{fig:masses-by-type} shows the mass distribution for all BHs bound to the Milky Way. The median mass of these BHs ($7.2 \msun$) is similar to the masses of BHs detected in XRBs \citep[e.g.,][]{Corral-Santana+2016}, but the full distribution extends up to ${\sim}55 \msun$.

We find that the vast majority ($91\%)$ of BHs are isolated at present day \citepalias[e.g.,][]{Olejak+2020:2020A&A...638A..94O}, despite having assumed that they are entirely formed in binaries. The left pie chart in Figure~\ref{fig:masses-by-type} shows the contribution of different types of BHs. Around two thirds of isolated BHs are the result of binary disruption after a core-collapse event ($59\%$ of the total population), while the remaining third are isolated as their progenitor binary merged before either companion reached core-collapse ($33\%$ of the total population) \citep{Sana+2012:2012Sci...337..444S}.

\begin{figure}
    \centering
    \includegraphics[width=\columnwidth]{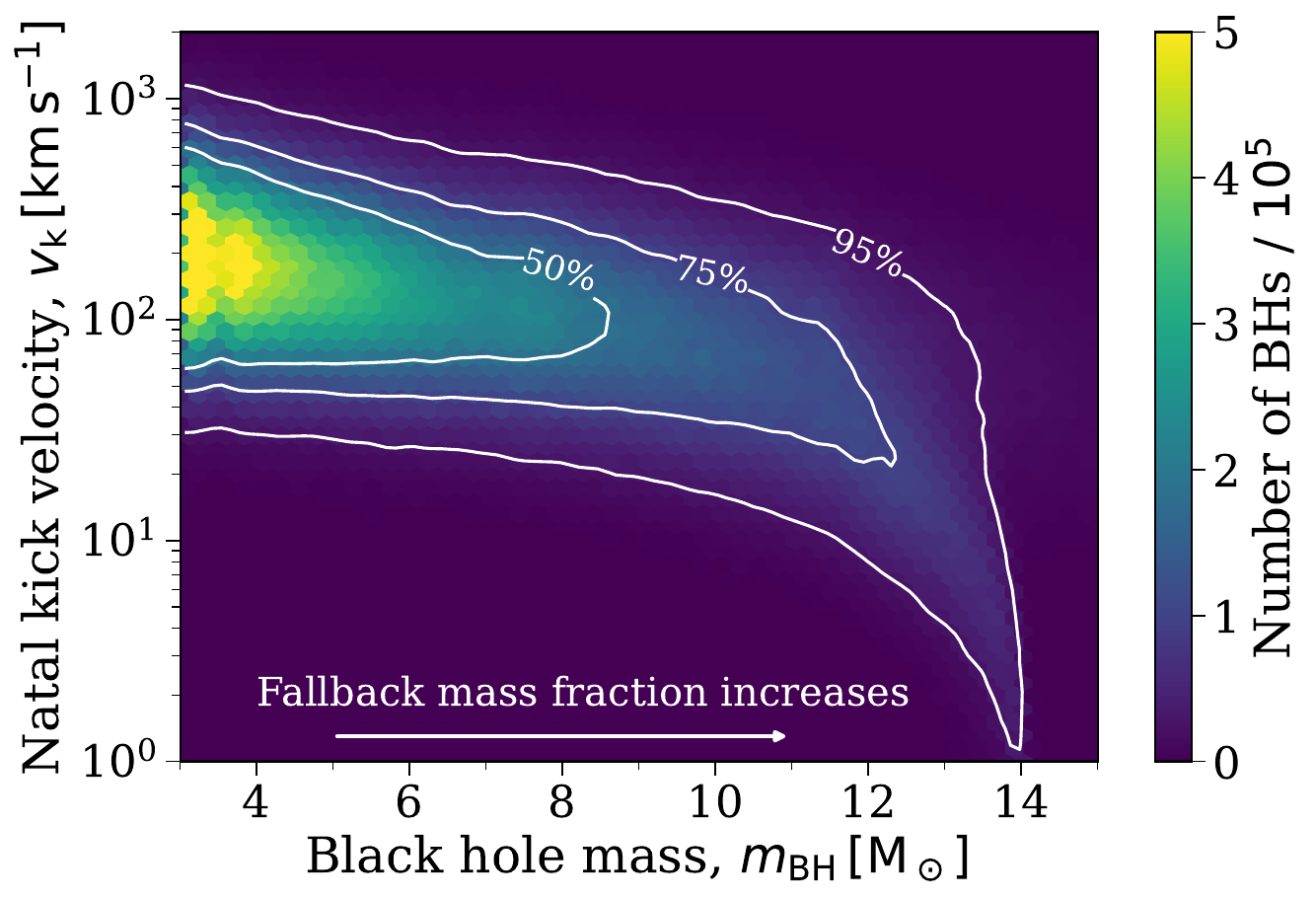}
    \caption{BH natal kicks decrease with increasing mass in the fiducial model \citep[applying][]{Fryer+2012:2012ApJ...749...91F}. This trend is driven by the increasing fallback mass fraction at higher masses, which negates the asymmetry in core-collapse events that produces kicks. The relation is shown with a 2D density distribution in BH mass and natal kick velocity. The white contour lines show the 50th, 75th, and 95th percentile lines.}
    \label{fig:kicks-vs-mass}
\end{figure}

The mass distribution of BHs is dependent on their formation channel due to the mass dependence of natal kicks. In Figure~\ref{fig:masses-by-type}, we plot separate histograms for BHs that are isolated due to a disruption (purple), isolated as a result of forming after a stellar merger (orange), and in a bound binary at present day (red). At low masses ($\leq 10 \msun$), the distribution is dominated by isolated BHs, with almost no contribution from BHs in binaries. However, at higher masses, most BHs are in bound binaries or from stellar mergers. This preference for high-mass BHs is also notable in the rightmost pie chart in Figure~\ref{fig:masses-by-type}, which shows each channel's contribution to the total mass of BHs (compared to the left pie chart, which shows the total \textit{number}).

BHs in binaries are favoured at high masses due to the mass-dependence of BH natal kicks in our simulations. After a core-collapse event, some fraction of the ejecta mass may fall back onto the collapsing core, negating any asymmetry (and therefore, natal kick) associated with it. We calculate the fallback mass fraction following \citet{Fryer+2012:2012ApJ...749...91F}, and scale natal kicks by the fraction of ejecta mass that falls back. In Figure~\ref{fig:kicks-vs-mass}, we show how the natal kick applied to BHs is typically lower for higher-mass BHs. In particular, low-mass BHs ($\leq 10 \msun$) receive natal kicks similar to those for NSs. However, as BH masses increases, a larger fraction of mass falls back, damping natal kicks. For BHs with masses $\gtrsim 14 \msun$, natal kicks from baryons are almost entirely negated. The reduction of natal kicks above this mass allows BH progenitors to avoid disrupting their binaries during their core collapse. For this reason, at large masses (shown in the inset panel) BHs in binaries dominate those from disrupted systems. BHs from stellar mergers are also more prominent at high masses, because these mergers increase the mass budget available for BH formation.

\subsection{Spatial distributions}\label{sec:fid_spatial}

\begin{figure*}[tb]
    \centering
    \includegraphics[width=\textwidth]{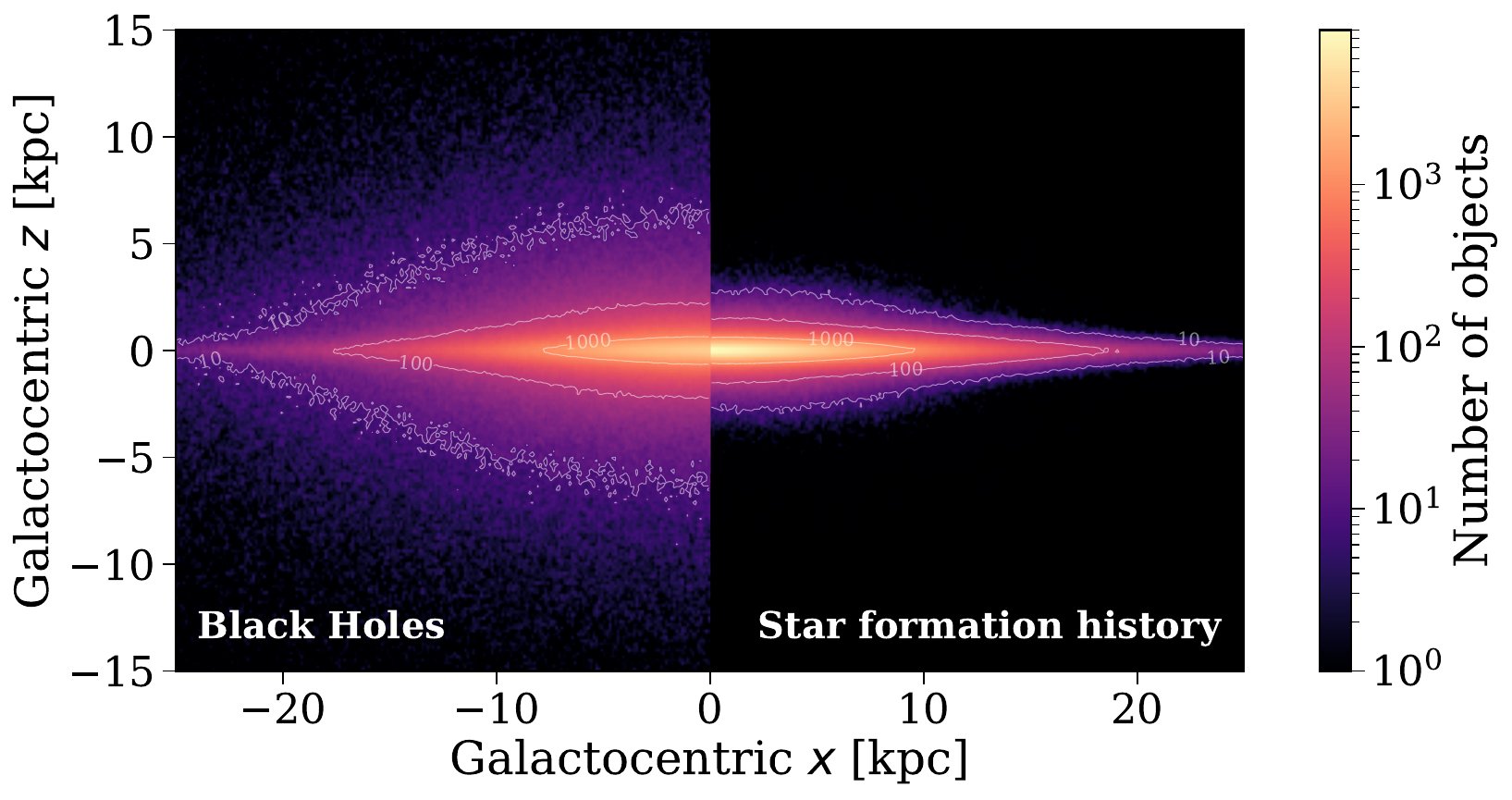}
    \caption{BHs are more diffusely distributed in the Galaxy than visible stars as a result of natal kicks. The left side shows the 2D density distribution of simulated BHs. The right side shows the distribution of an equal number of stars following the assumed star formation history model.}
    \label{fig:side-on-density}
\end{figure*}

Natal kicks applied to BHs significantly alter their spatial distributions compared to the visible galaxy. This contrast can be seen in Figure~\ref{fig:side-on-density}, where we show an edge-on view of the simulated Galaxy. The left half shows the simulated population of BHs, and the right half shows an equal number of stars drawn from the assumed star formation history.

\begin{figure}
    \centering
    \includegraphics[width=\columnwidth]{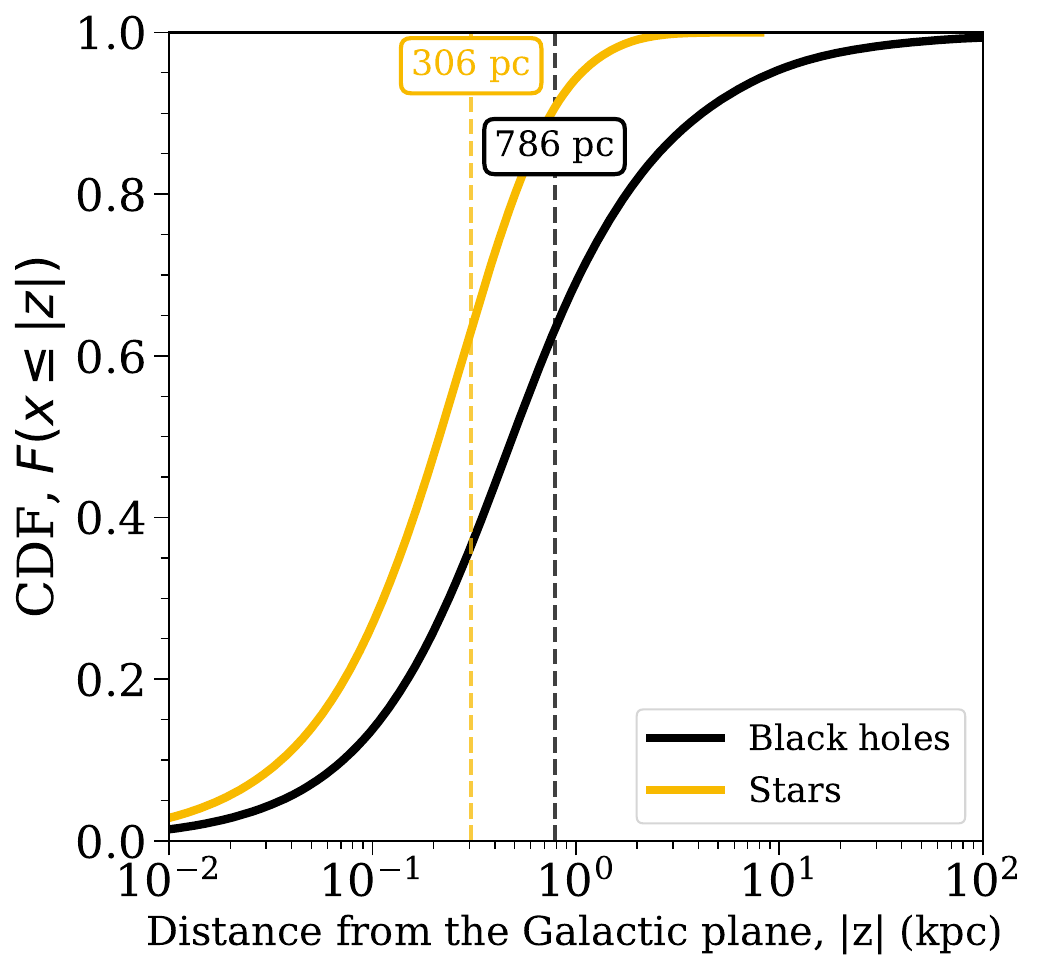}
    \caption{The scale height of Galactic BHs is increased relative to the visible galaxy due to supernova natal kicks. The cumulative distribution functions of absolute Galactic heights are shown in solid lines (stars in gold and BHs in black). Dashed lines are shown at the height that contains $1 - 1/e$ of the distribution (equivalent to the scale height of an exponential distribution). The stellar distribution is drawn from the star formation history and thus is representative of all stars.}
    \label{fig:scale-heights}
\end{figure}

The most obvious difference in Figure~\ref{fig:side-on-density} is the significantly larger scale height of the BHs compared to the visible stars. We quantify this in Figure~\ref{fig:scale-heights}, where we plot the marginal distributions of the distance from the Galactic plane as cumulative distribution functions (CDFs), with stars in gold and BHs in black. The vertical distribution of stars follows an exponential (as prescribed by the star formation history), but the BHs have an extended tail to large heights, as a result of the additional heating of Galactic orbits from natal kicks. 

We calculate the effective scale height $h_{z, {\rm eff}}$ of each distribution as the height at which the CDF reaches $1 - 1/e$, which would recover the scale height $h_z$ for an exponential distribution. For the visible stars, $h_{z, {\rm eff}}=306 \unit{pc}$, while for the BHs we find $h_{z, {\rm eff}}=786 \unit{pc}$, which is around $2.5\times$ larger due to natal kicks. This scale height is more comparable with that of the thick, high $[\alpha / \rm Fe]$, disk \citep[$500$--$1000 \unit{pc}$, e.g.,][Figure 5]{Bovy+2012:2012ApJ...753..148B}.

The difference in BH kinematics is primarily driven by natal kicks because their ejection velocities from disrupted binaries are typically low. Even in the absence of a natal kick, a BH may be ejected from a binary by a Blaauw kick \citep{Blaauw+1961}, which comes from symmetric mass loss, or the kick of a companion. In this case, its ejection velocity closely follows its progenitor's pre-supernova orbital velocity \citep[e.g.,][]{Wagg+2025:2025OJAp....8E..85W}. Similarly, when a BH natal kick disrupts a binary, the orbital velocity also contributes to the BH's ejection velocity \citep[e.g.,][Appendix A]{Pfahl+2002}. However, we find that, for BH progenitors, this orbital velocity is often very low (typically $10 \unit{km}{s^{-1}}$, but up to $60 \unit{km}{s^{-1}}$) compared to the natal kick and thus has a negligible impact on the kinematics of BHs.

Beyond the larger-scale spatial distribution of BHs, we also investigate their distribution in the solar neighbourhood. We estimate the number density close to the Sun by totalling the occurrence of BHs in a torus with a major radius of $8 \unit{kpc}$ and a minor radius of $0.5 \unit{kpc}$. In this region, we find that the number density of BHs is $\num{1.6e-4} \unit{pc^{-3}}$. Therefore, based on this number density we estimate that the nearest BH to the Sun is, on average, $19 \unit{pc}$ away.

\subsection{BHs further from the Galactic plane are typically less massive}\label{sec:correlations}

Natal kicks, which alter the spatial distribution of BHs, have different typical magnitudes depending on the mass of the BH (as we discussed in Section~\ref{sec:fid_rates_binarity}). Figure~\ref{fig:kicks-vs-mass} highlights how an increase in BH mass can drastically reduce a natal kick as a result of ejecta mass falling back onto the collapsing core. A natural outcome of this trend is that less-massive BHs, which receive larger kicks, should be more spatially distinct from the visible galaxy. As a result, the BH mass distribution should be skewed to lower masses at large distances from the plane.

\begin{figure}
    \centering
    \includegraphics[width=\columnwidth]{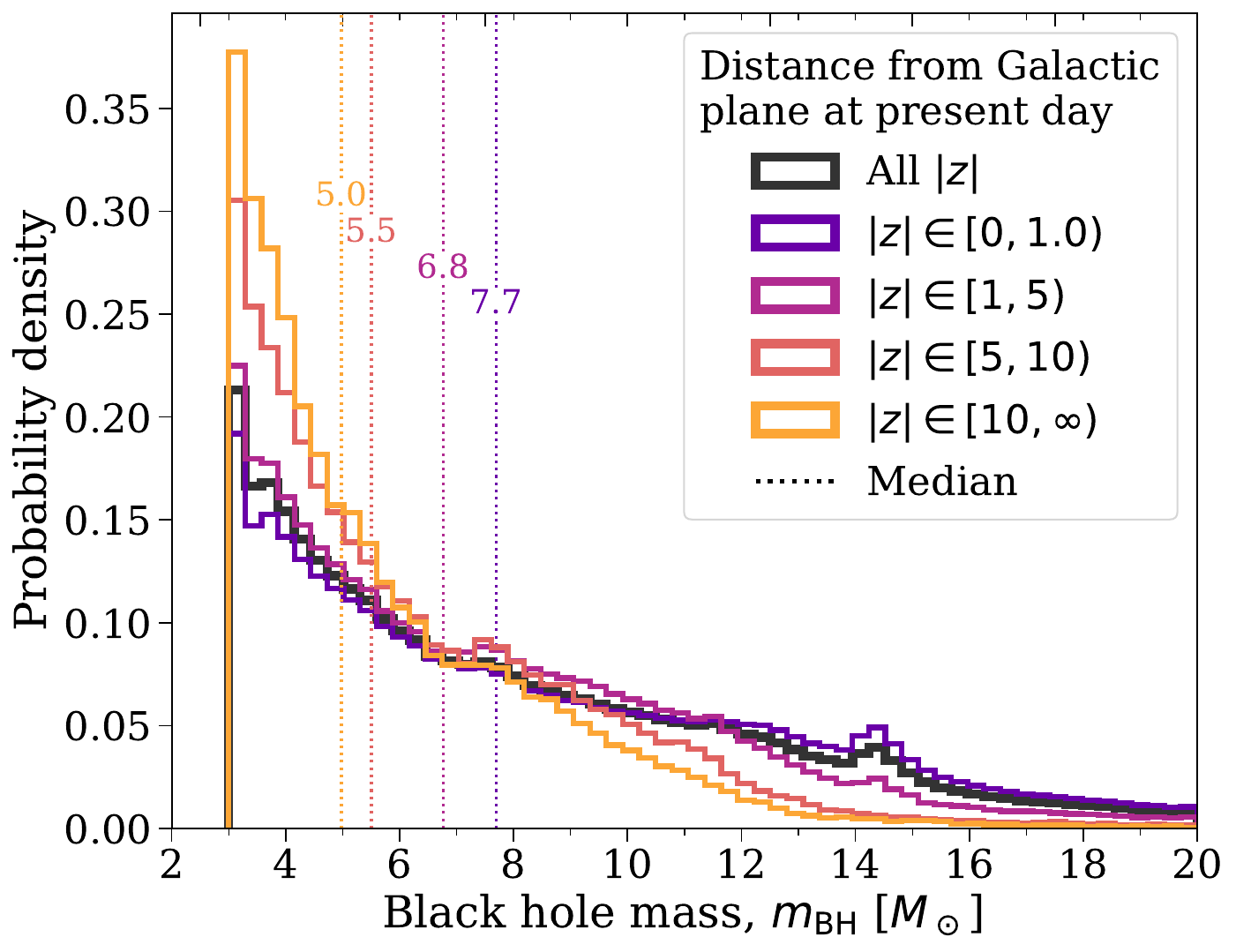}
    \caption{BHs that are further from the Galactic plane are less massive on average at present day, due to low-mass BHs receiving large natal kicks on average. Normalised histograms of BH mass, separated into bins of distance from the Galactic plane at present day, $|z|$. The black histogram shows the total distribution as a guide. Dotted lines indicate the median value of each histogram.}
    \label{fig:average-mass-by-z}
\end{figure}

Our simulations confirm that there are strong correlations between BH mass and present-day $\abs{z}$ position, as shown in Figure~\ref{fig:average-mass-by-z}. We plot normalised histograms of BH mass, separated into bins of $|z|$, with dotted lines to indicate the median of the distribution. Close to the plane ($|z| < 1 \unit{kpc}$), the median BH mass is higher, at around $7.7 \msun$ (dark purple histogram). In this region, low-mass BHs are less prevalent as they have received stronger natal kicks that propel them onto more heated Galactic orbits. With increasing distance from the plane, the distributions therefore skew to lower masses, such that BHs with $\abs{z} \ge 10\unit{kpc}$ have a median mass of $5\msun$ (orange histogram). We tabulate summary statistics for different kinematic populations in Table~\ref{tab:single_bh_masses}.

\begin{table}
    \centering
    \begin{tabular}{l c c c}
      \hline
      Population & \multicolumn{3}{c}{BH mass percentiles [M$_\odot$]} \\
      & \eqcol{25th} & \eqcol{50th} & \eqcol{75th} \\
      \hline
      Bound to Milky Way & \eqcol{4.6} & \eqcol{7.2} & \eqcol{11.3} \\
      \quad $\hookrightarrow \abs{z} < 1 \unit{kpc}$ & \eqcol{4.8} & \eqcol{7.7} & \eqcol{12.2} \\
      \quad $\hookrightarrow \abs{z} \ge 1 \unit{kpc}$ & \eqcol{4.2} & \eqcol{6.2} & \eqcol{9.4} \\
      Escaped Milky Way & \eqcol{3.5} & \eqcol{4.4} & \eqcol{6.0} \\
      \hline
    \end{tabular}
    \caption{Less-massive BHs are more often displaced from the midplane and are often ejected from the Galaxy due to the mass dependence of natal kicks. This table gives the median BH mass and interquartile range, for different kinematic populations.}
    \label{tab:single_bh_masses}
\end{table}

\subsection{Velocity distributions}\label{sec:fid_vels}

BHs are kinematically hotter than the stellar population as a result of the same natal kicks that have dispersed them around the Galaxy. We find that the velocity dispersion of Milky Way BHs is 50\% larger than the stellar velocity dispersions in the radial, $\sigma_R$, and vertical, $\sigma_z$, directions (see Figure~\ref{fig:vel_components}). 

The dispersion in the azimuthal direction, $\sigma_\phi$, is 70\% larger for BHs than for stars, which is a bigger difference than for $\sigma_R$ or $\sigma_z$. The larger difference in this case is driven by the fact that natal kicks can drive BHs onto retrograde Galactic orbits, skewing the distribution of the azimuthal Galactocentric velocity, $v_\phi$.
Around 4\% of BHs have retrograde orbits ($v_\phi < 0$) as a result of strong natal kicks (see Figure~\ref{fig:vel_Lz_E}). A subset of these retain disc-like orbits, but in the opposite direction from the stellar distribution. BHs with retrograde orbits are typically much less massive, with a median mass of $4.7\msun$ compared to $7.3\msun$ for prograde orbits. This mass trend is present for the same reason as the correlation between BH mass and distance from the plane: low-mass BHs have lower fallback mass fractions, and thus larger natal kicks.

\subsection{Free-floating BHs that escape the Galaxy}\label{sec:fid_free_floating}

A small fraction of BHs manage to escape the gravitational potential of the Milky Way. We find that around $3\%$ of BHs have positive Galactic orbital energies (see Figure~\ref{fig:vel_Lz_E}) and therefore escape the Galaxy. Of these escaped BHs, the vast majority are isolated as a result of the strong natal kick required to leave the Galaxy, such that only $0.03\%$ of them are ejected with a companion in tow.

Due to the mass-dependence of natal kicks, the escapees are typically the lowest-mass BHs, with a median mass of $4.4\msun$. Since only a small fraction of BHs escape, this does not shift the mass distribution of Milky Way BHs significantly. However, in dwarf galaxies, one may expect that the mass distribution of BHs skews to higher mass BHs as a result of the ejection of low-mass BHs. Overall, the exodus of BHs from the Milky Way corresponds to a total Galactic mass loss of $2.4 \times 10^7 \msun$ over $12 \unit{Gyr}$. Adopting the median ejected BH mass, this corresponds to the Galaxy losing a BH roughly once every two millennia.

\section{Sensitivity to model variations}\label{sec:variations}

\begin{figure}
    \centering
    \includegraphics[width=\columnwidth]{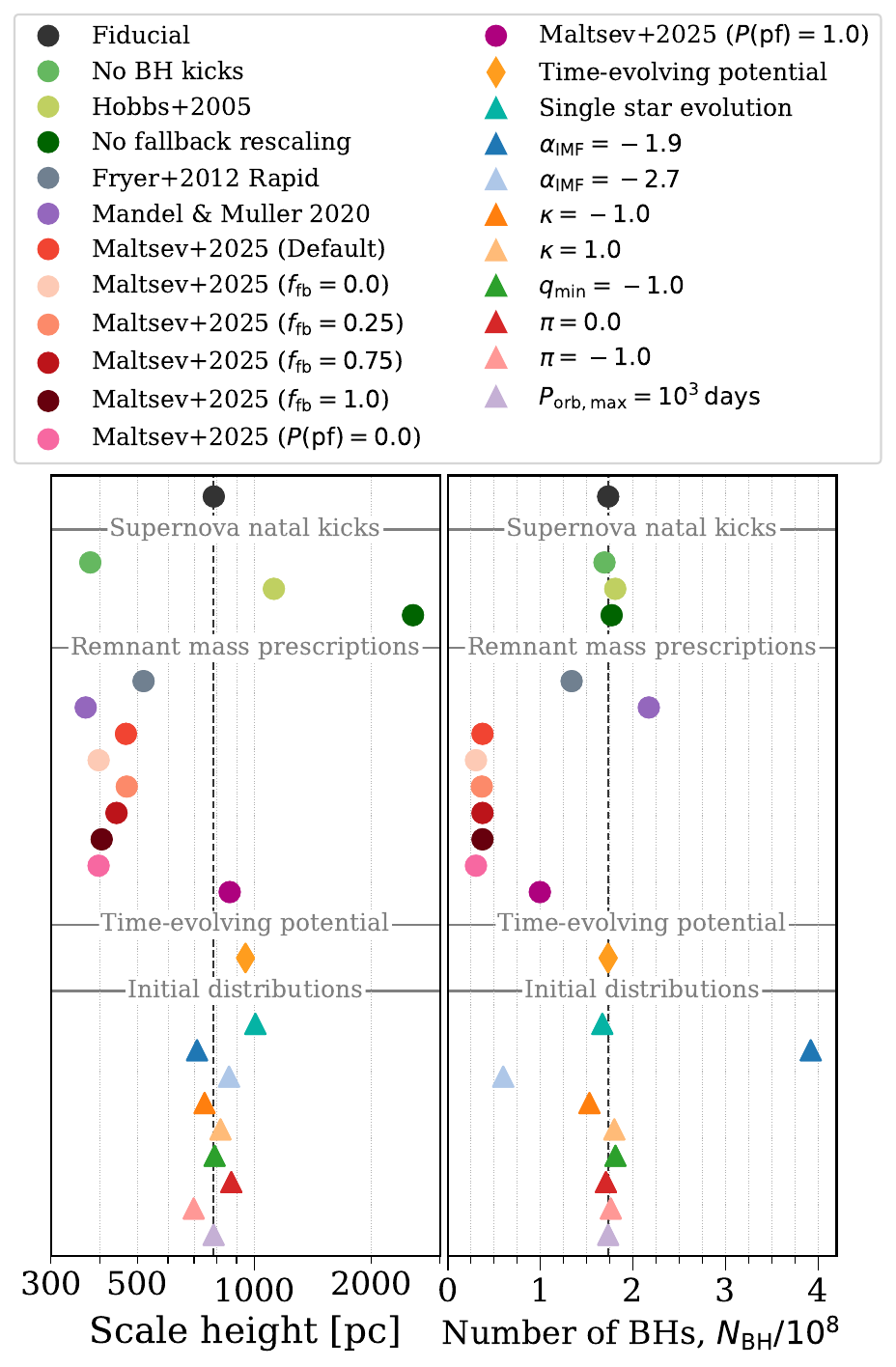}
    \caption{\textbf{Left:} The scale height of BHs in our model variations. The dashed line and shaded area indicates the scale height for the fiducial model. \textbf{Right:} The number of BHs in the Milky Way at present day. The dashed line indicates the total for the fiducial model. Variations of natal kicks and remnant mass prescriptions are shown as circles, while variations of initial distributions are triangles. See Figures~\ref{fig:summary-stats}~\&~\ref{fig:summary-other-cols}.}
    \label{fig:summary-stats}
\end{figure}

In this Section, we evaluate the sensitivity of our results to variations in binary physics, initial binary distribution, and the Galactic potential. We focus on variations that are most likely to impact BH formation rates and spatial distributions. In Table~\ref{tab:summary_table}, we present a series of summary statistics for each of the variations, which are then visualized in Figure~\ref{fig:summary-stats} for the scale height and BH production rate; visualizations of the other parameters are included in Figure~\ref{fig:summary-other-cols}. We explore the most important of these parameter variations in the subsections below.

\begin{table*}
    \centering
    \begin{tabular}{lcccccccc}
\toprule
Model & $N_{\rm BH} / 10^8$ & $M_{\rm BH} / 10^9 M_\odot$ & $h_{z, {\rm eff}}$ (pc) & $f_{\rm esc}$ & $\median{m_{|z| < 1 \, {\rm kpc}}}$ & $\median{m_{|z| \geq 1 \, {\rm kpc}}}$ & $f_{\rm bound}$ & $f_{\rm merger}$ \\
\midrule
\textbf{Fiducial} & $1.73$ & $1.50$ & $786$ & 2.7\% & $7.70$ & $6.22$ & $9\%$ & $33\%$ \\
\addlinespace[0.5em]
\multicolumn{8}{l}{\textbf{Supernova natal kicks}} \\
\quad No BH kicks & $1.69$ & $1.46$ & $379$ & 0.0\% & $6.84$ & $8.62$ & $33\%$ & $35\%$ \\
\quad Hobbs+2005 & $1.81$ & $1.56$ & $1122$ & 2.6\% & $8.15$ & $6.08$ & $7\%$ & $31\%$ \\
\quad No fallback rescaling & $1.77$ & $1.54$ & $2557$ & 15.5\% & $6.38$ & $7.75$ & $0\%$ & $30\%$ \\
\addlinespace[0.5em]
\multicolumn{8}{l}{\textbf{Remnant mass prescriptions}} \\
\quad Fryer+2012 Rapid & $1.34$ & $1.50$ & $519$ & 0.7\% & $9.30$ & $8.62$ & $15\%$ & $34\%$ \\
\quad Mandel\&Muller2020 & $2.17$ & $1.90$ & $368$ & 0.1\% & $8.19$ & $7.11$ & $26\%$ & $41\%$ \\
\quad Maltsev+2025 & $0.38$ & $0.47$ & $468$ & 0.8\% & $9.66$ & $7.83$ & $18\%$ & $31\%$ \\
\multicolumn{8}{l}{\textit{\quad Maltsev+2025 fallback mass fraction}} \\
\quad\quad $f_{\rm fb} = 0.0$ & $0.30$ & $0.42$ & $398$ & 0.3\% & $9.82$ & $9.92$ & $21\%$ & $31\%$ \\
\quad\quad $f_{\rm fb} = 0.25$ & $0.37$ & $0.45$ & $470$ & 1.7\% & $9.70$ & $5.19$ & $18\%$ & $30\%$ \\
\quad\quad $f_{\rm fb} = 0.75$ & $0.38$ & $0.49$ & $442$ & 0.3\% & $9.76$ & $9.60$ & $18\%$ & $31\%$ \\
\quad\quad $f_{\rm fb} = 1.0$ & $0.38$ & $0.51$ & $405$ & 0.3\% & $10.01$ & $10.60$ & $21\%$ & $31\%$ \\
\multicolumn{8}{l}{\textit{\quad Maltsev+2025 partial fallback probability}} \\
\quad\quad $P({\rm pf}) = 0.0$ & $0.30$ & $0.42$ & $398$ & 0.3\% & $9.82$ & $9.93$ & $21\%$ & $31\%$ \\
\quad\quad $P({\rm pf}) = 1.0$ & $0.99$ & $0.87$ & $864$ & 2.1\% & $7.51$ & $6.54$ & $8\%$ & $33\%$ \\
\addlinespace[0.5em]
\multicolumn{8}{l}{\textbf{Galactic gravitational potential}} \\
\quad Time-evolving & $1.73$ & $1.50$ & $949$ & 6.8\% & $7.78$ & $6.55$ & $9\%$ & $33\%$ \\
\addlinespace[0.5em]
\multicolumn{8}{l}{\textbf{Single star evolution}} \\
\quad $f_{\rm bin} = 0.0$ & $1.67$ & $1.46$ & $1006$ & 3.4\% & $8.55$ & $6.45$ & - & - \\
\addlinespace[0.5em]
\multicolumn{8}{l}{\textbf{Initial mass function}} \\
\quad $\alpha_{\rm IMF} = -1.9$ & $3.92$ & $3.70$ & $712$ & 2.3\% & $8.43$ & $6.50$ & $12\%$ & $31\%$ \\
\quad $\alpha_{\rm IMF} = -2.7$ & $0.60$ & $0.48$ & $861$ & 3.2\% & $7.10$ & $5.98$ & $6\%$ & $34\%$ \\
\addlinespace[0.5em]
\multicolumn{8}{l}{\textbf{Initial mass ratio distribution}} \\
\quad $\kappa = -1.0$ & $1.53$ & $1.33$ & $745$ & 2.6\% & $7.89$ & $6.41$ & $8\%$ & $56\%$ \\
\quad $\kappa = 1.0$ & $1.80$ & $1.56$ & $818$ & 2.8\% & $7.71$ & $6.19$ & $10\%$ & $25\%$ \\
\quad $q_{\rm min} = -1.0$ & $1.81$ & $1.57$ & $790$ & 2.7\% & $7.68$ & $6.20$ & $9\%$ & $30\%$ \\
\addlinespace[0.5em]
\multicolumn{8}{l}{\textbf{Initial orbital period distribution}} \\
\quad $\pi = 0.0$ & $1.71$ & $1.48$ & $872$ & 3.0\% & $7.90$ & $6.17$ & $9\%$ & $23\%$ \\
\quad $\pi = -1.0$ & $1.76$ & $1.53$ & $698$ & 2.4\% & $7.51$ & $6.33$ & $8\%$ & $43\%$ \\
\quad $P_{\rm orb, max} = 10^3\,{\rm days}$ & $1.73$ & $1.50$ & $786$ & 2.7\% & $7.71$ & $6.22$ & $9\%$ & $33\%$ \\
\hline\hline
\textbf{Summary statistics} \\
\quad Minimum & $0.30$ & $0.42$ & $368$ & $0.0\%$ & $6.38$ & $5.19$ & $0\%$ & $23\%$ \\
\quad Median & $1.69$ & $1.48$ & $745$ & $2.4\%$ & $7.90$ & $6.50$ & $10\%$ & $31\%$ \\
\quad Maximum & $3.92$ & $3.70$ & $2557$ & $15.5\%$ & $10.01$ & $10.60$ & $33\%$ & $56\%$ \\
\bottomrule
\end{tabular}
    \caption{Summary statistics for the suite of simulations, described in Section~\ref{sec:variations}. $N_{\rm BH}$ is the number of BHs produced over the Milky Way's history, $M_{\rm BH}$ is the total mass containing in those BHs, $h_{z, {\rm eff}}$ is the BH scale height, $f_{\rm esc}$ is the fraction of BHs that escaped the Milky Way potential, $\median{m_{|z| < 1}}$ and $\median{m_{|z| \ge 1}}$ are the average BH masses close to and far from the Galactic plane respectively, $f_{\rm bound}$ is the fraction of BHs that remain in bound binaries at present day, and $f_{\rm merger}$ is the fraction of BHs formed from a merger product progenitor.}
    \label{tab:summary_table}
\end{table*}

\subsection{Milky Way BH kinematics can constrain natal kick models}\label{sec:kick-variations}

Observationally and theoretically, the natal kick distribution of BHs remains highly uncertain. Studies of pulsar kinematics have long demonstrated that at least some neutron stars must receive strong natal kicks \citep[e.g.,][]{Lyne+1994}. However, it is less well understood how these distributions may be applied to BHs \citep[e.g.,][]{Janka+2017:2017ApJ...837...84J, Repetto+2012:2012MNRAS.425.2799R, atri:19, Burrows+2025:2025ApJ...987..164B}. BHs have almost exclusively been observed in binaries, biasing results to lower natal kicks that are less likely to disrupt binaries. Observations of some BHs in binaries imply the BH must have received almost no natal kick \citep[e.g.\ VFTS 473,][]{Vigna-Gomez+2024:2024PhRvL.132s1403V}, whilst others require the natal kick to be strong \citep[e.g.\ Swift J1727.8-162,][]{MataSanchez+2025:2025A&A...693A.129M}. Supporting these observations, it has been shown that data from \gaia DR3 are consistent with a scenario in which some BHs form via direct collapse and receive no kick, whilst others explode in SNe and receive strong kicks \citep{Nagarajan+2025:2025PASP..137c4203N}. A common method for modelling this in population synthesis is to scale BH kicks by their fallback mass fraction \citep{Fryer+2012:2012ApJ...749...91F}, such that full fallback BHs receive no kick, as we have demonstrated in Figure~\ref{fig:kicks-vs-mass}.

Given the uncertainty on BH natal kicks, we re-run the \cogsworth simulations for representative bounding cases. In the fiducial model, we adopt the \citet{Disberg+2025:2025ApJ...989L...8D} distribution for CCSN natal kicks, and scale BH kicks by their fallback mass fraction following \citet{Fryer+2012:2012ApJ...749...91F}. For the bounding variations, we repeat the fiducial simulation assuming (a) BHs never receive any natal kick (regardless of mass) and (b) BHs receive natal kicks following the same distribution as NSs from CCSN. For historical reasons, we include an additional variation assuming the \citet{Hobbs+2005} distribution for CCSN natal kicks (and still apply fallback scaling). This distribution has been shown to not be statistically representative of the sample \citep{Disberg+2025:2025ApJ...989L...8D}, but we include it to facilitate comparison to earlier work.

The scale height of BHs scales directly with the average strength of assumed natal kicks. Figure~\ref{fig:summary-stats} shows that the scale height of BHs decreases significantly in the absence of natal kicks, from $h_{z, {\rm eff}} = 790\unit{pc}$ to $380\unit{pc}$. This value is still higher than the scale height of visible stars for two reasons. Firstly, BHs are preferentially formed in the kinematically hotter thick disc since it has a lower average metallicity, which favours BH formation. Second, BHs can still be ejected from binaries by a NS companion's natal kick. When assuming that BHs receive the same kicks as NSs (i.e.\ without applying fallback scaling), the scale height increases to $2570\unit{pc}$, the largest of any variation we consider. This model also increases the fraction of BHs that escape the Milky Way to $16\%$ (see Table~\ref{tab:summary_table}).

\begin{figure}
    \centering
    \includegraphics[width=\columnwidth]{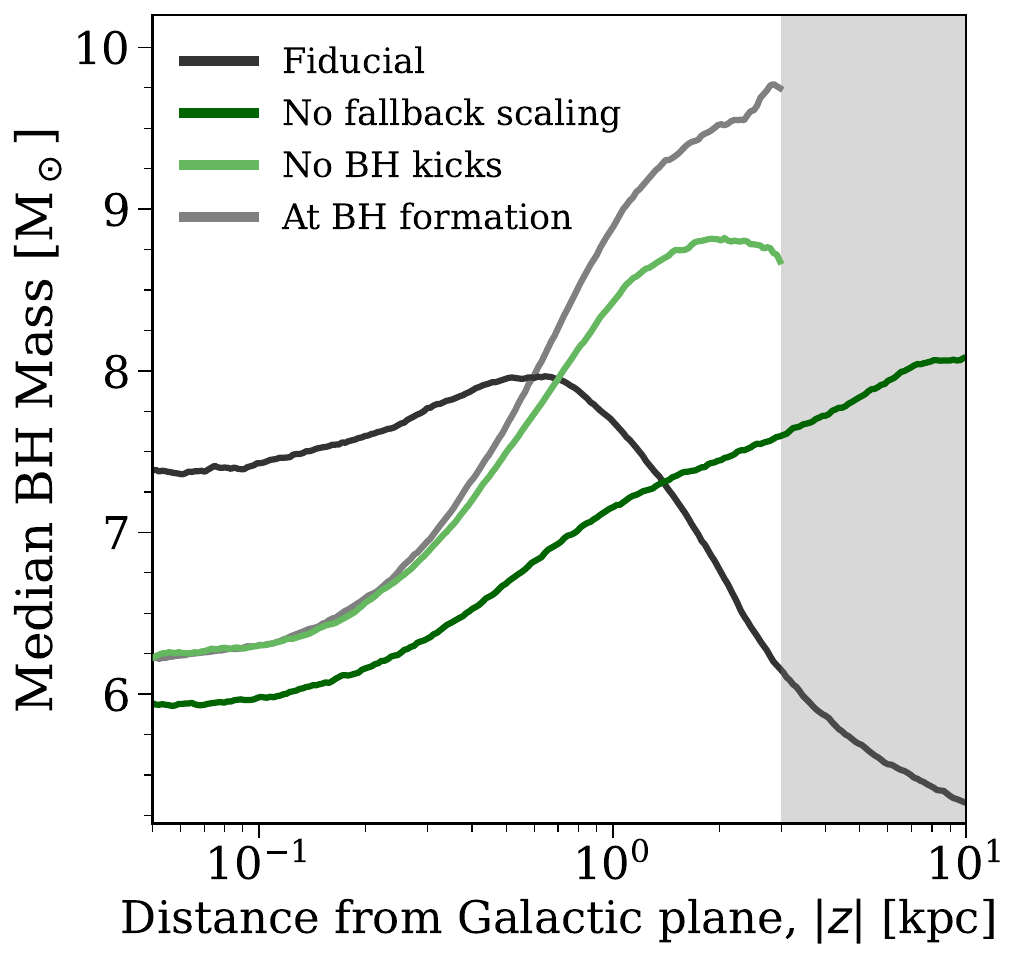}
    \caption{Correlations between BH mass and location are sensitive to choices regarding BH natal kicks. Lines show a running median of BH mass (averaged over a window of 0.25 dex) as a function of distance from the Galactic plane. The black line shows the running average when using the present-day location of BHs in the fiducial model (this is an alternate visualisation of the same trend as Figure~\ref{fig:average-mass-by-z}). The grey line is the same but using the location at which the BHs \textit{form}. The green lines show our bounding variations for natal kicks (light green assumes BH receive no kicks, dark green assumes BH receive the same kicks as NSs). The grey area beyond $3 \, {\rm kpc}$ indicates the region that is poorly populated by star formation, and as such the distributions at formation (grey) and without BH kicks (light green) are not plotted here.}
    \label{fig:average-mass-by-z-kicks}
\end{figure}

The correlation between BH mass and distance from the Galactic plane is also affected by varying how natal kicks are applied to BHs. In Figure~\ref{fig:average-mass-by-z-kicks}, we show how the median BH mass changes as a function of distance from the plane for each of the natal kick variations. The fiducial model shows the same trend we discussed in Figure~\ref{fig:average-mass-by-z}: as you move further from the Galactic plane, BHs are typically lower mass as a result of fallback-scaled natal kicks.

It is notable that the present-day correlation between BH mass and height is entirely opposite of that imprinted when the BHs form, which is shown as the grey line in Figure~\ref{fig:average-mass-by-z-kicks}. At formation, BHs further from the plane are \textit{more} massive on average. This trend is driven by the lower stellar metallicity of the thick disc, which contributes more to BH formation at large $\abs{z}$. The lower metallicity leads to weaker winds, allowing the eventual BH to have retained more of its initial stellar mass. However, low-mass BHs receive large kicks, and this is sufficient to invert the correlation by present day.

As one would expect, removing BH natal kicks means that they retain nearly the same correlation between average mass and $|z|$ as was imprinted at formation. The slight deviation at larger values of $|z|$ is driven by NS companions to BHs. NSs still receive natal kicks in this variation, and as such can disrupt binaries, ejecting BHs with their pre-SN orbital velocities (as discussed in Section~\ref{sec:fid_spatial}). NSs are most often companions to low-mass BHs and thus this additional orbital heating is only relevant to low-mass BHs, skewing the large $|z|$ median to lower masses. Moreover, the orbital velocities of these binaries are typically around $10 \unit{km}{s^{-1}}$ and not more than $60 \unit{km}{s^{-1}}$ (much lower than average natal kicks, see Figure~\ref{fig:kicks-vs-mass}) and as such only produce a weak deviation from the BH formation relation.

More interestingly, when removing the fallback scaling for BHs, the average BH mass also increases with increasing distance from the galactic plane, similar to the trend at formation. Since all BH kicks are drawn from the same distribution, massive BHs that are formed at larger heights are less bound to the Galaxy and the same natal kick propels them even further from the plane. Overall, observations of BH scale heights and masses offer potential constraints on BH natal kick models.


\subsection{The Milky Way BH mass distribution can calibrate remnant mass prescriptions}\label{sec:mass-from-rmp}

Rapid binary population synthesis codes rely on uncertain mappings from pre-CCSN core structures to the final remnant masses. These remnant mass prescriptions are based on both 3-dimensional and parameterised 1-dimensional SN simulations. Distilling an unambiguous relation between pre-SN stellar structure and remnant mass (and explosion outcome) is difficult \citep[e.g.,][]{Farmer+2016:2016ApJS..227...22F,Mandel+2020:2020MNRAS.499.3214M, Patton+2020:2020MNRAS.499.2803P, Myers+2026:2026arXiv260422605M, Renzo+2026:2026arXiv260621824R}, given the challenges in detailed modelling of simulated CCSN populations \citep[e.g.,][]{Janka+2012:2012ARNPS..62..407J,Muller+2020:2020LRCA....6....3M,Vartanyan+2022:2022MNRAS.510.4689V,Mezzacappa+2026:2026arXiv260424970M}, which remains a long-standing problem.

Given the absence of a strong astrophysical prior from theory, we explore the sensitivity of our results to the choice of remnant mass prescription, with the aim of using Milky Way BHs as potential calibrators. The fiducial model adopts the \citet{Fryer+2012:2012ApJ...749...91F} \textit{delayed} prescription, but there are a number of plausible alternatives that we consider.

\subsubsection{Alternative remnant mass prescriptions}

The first of these variations is the \textit{rapid} prescription from \citet{Fryer+2012:2012ApJ...749...91F}, which requires successful explosions to occur rapidly (within 250 ms) after core bounce. We also implement the probabilistic model from \citet{Mandel+2020:2020MNRAS.499.3214M}, based on updated SN simulations. For this variation, we match a number of \cosmic parameters to the \citet{Mandel+2020:2020MNRAS.499.3214M} model choices, by setting the maximum neutron star mass to $2 \msun$, adopting the natal kick prescription from \citet{Mandel+2020:2020MNRAS.499.3214M}, and limiting baryonic mass loss to $0.1 \msun$ (in the fiducial model these settings are $3\msun$, the \citealt{Disberg+2025:2025ApJ...989L...8D} prescription, and $0.5\msun$ respectively). 

We also evaluate a more recent remnant mass prescription presented in \citet{Maltsev+2025:2025AA...700A..20M} and \citet{Willcox+2025:2025arXiv251007573W}, based on semi-analytic supernova models \citep{Muller+2016:2016PASA...33...48M} and stellar simulations \citep{Schneider+2021:2021A&A...645A...5S, Schneider+2023:2023ApJ...950L...9S}. This prescription accounts for the impact of mass transfer on the core structure of a star. It specifies that BH formation is non-monotonic with progenitor mass, leading to ``islands of explodability''\footnote{See also \citet{Patton+2020:2020MNRAS.499.2803P}, which also accounts for islands of explodability, but is not implemented in \cosmic.}. As such, certain mass ranges in this prescription stochastically produce BHs with partial fallback with a probability $P({\rm pf})$, which is assumed to be $10\%$ by default, and otherwise produce NSs. In this mass range, the fallback fraction is assumed to take on a fixed value of $f_{\rm fb}$, which is assumed to be $0.5$ by default. We explore several variations using this prescription, varying both $f_{\rm fb}$ and $P({\rm pf})$ between 0 and 1 (see Table~\ref{tab:summary_table} and Figure~\ref{fig:summary-stats}).

\begin{figure}
    \centering
    \includegraphics[width=\columnwidth]{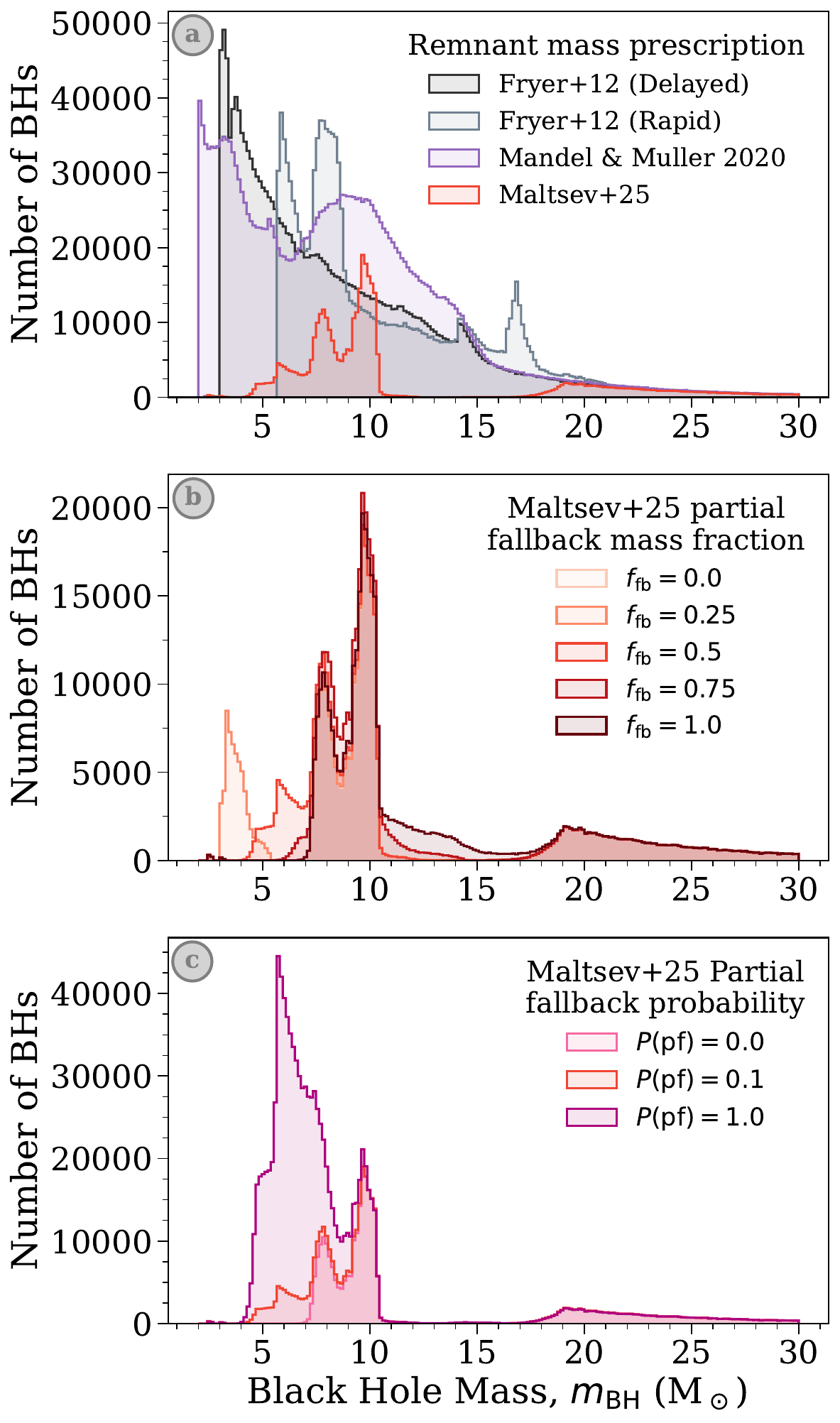}
    \caption{The distribution, and number, of BHs is very sensitive to variations in the remnant mass prescription. Each panel shows the distribution of BH masses for a range of variations. Panel \textbf{(a)} changes the remnant mass prescription, where \citet{Maltsev+2025:2025AA...700A..20M} has the default values of $f_{\rm fb} = 0.5$ and $P(\rm pf) = 0.1$. Panel \textbf{(b)} varies the fallback mass fraction in the case of partial fallback for the \citet{Maltsev+2025:2025AA...700A..20M} prescription. Panel \textbf{(c)} varies the probability of partial fallback occurring in partial fallback ranges in the \citet{Maltsev+2025:2025AA...700A..20M} prescription. We additionally show cumulative distribution functions for BH mass in Figure~\ref{fig:rmp-cdfs}.}
    \label{fig:bh_mass_rmp}
\end{figure}

\subsubsection{Impact of varying the remnant mass prescription on the BH mass distribution}\label{sec:rmp_effect_on_bh_mass}

We re-run the \cogsworth simulations with the remnant mass prescriptions above, and show how the resulting BH mass distributions for each variation in Figure~\ref{fig:bh_mass_rmp}. As anticipated, the Milky Way BH mass distribution is highly sensitive to the choice of remnant mass prescription. All prescriptions produce distinct distributions, with no self-similarity in amplitude or shape. The prescriptions also lead to substantial differences in the total number of BHs, as can be seen in the right panel of Figure~\ref{fig:summary-stats} and Table~\ref{tab:summary_table}.  

\paragraph{\citet{Fryer+2012:2012ApJ...749...91F} \textit{rapid}} The upper panel of Figure~\ref{fig:bh_mass_rmp} shows that moving from the fiducial \textit{delayed} to the \textit{rapid} prescription from \citet{Fryer+2012:2012ApJ...749...91F} shifts the minimum BH mass from $3\msun$ to around $5.6\msun$ due to increased sensitivity to the sharp differences in core structure. Notably this prescription predicts two strong peaks, similar to \citet{Maltsev+2025:2025AA...700A..20M}, but at lower masses (around 6 and 8 \msun). Moreover, this model produces an overdensity of BHs around 17 \msun, likely due to the non-monotonicity of the remnant mass function in the \textit{rapid} model \citep[see Figure 11 of][]{Fryer+2012:2012ApJ...749...91F}.

\paragraph{\citet{Mandel+2020:2020MNRAS.499.3214M}} Compared to \citet{Fryer+2012:2012ApJ...749...91F} \textit{rapid}, the \citet{Mandel+2020:2020MNRAS.499.3214M} BH mass distribution is much smoother, due to the stochastic nature of the prescription smoothing sharp features. This model predicts one of the highest rates of BH formation, which is mostly driven by the lower maximum NS mass of $2 \msun$ (compared to $3 \msun$ in the other prescriptions), which allows for more BHs to be formed from lower mass stars.

\paragraph{\citet{Maltsev+2025:2025AA...700A..20M}}
In stark contrast to the other models, the \citet{Maltsev+2025:2025AA...700A..20M} prescription predicts a much lower rate of BHs (around $5\times$ lower, see Figures~\ref{fig:summary-stats}~\&~\ref{fig:rmp-cdfs}), and a more restricted range of masses. This decrease is because the prescription much more often forms NSs than others we have tested, and thus the paucity of BHs is balanced by an abundance of NSs. The default choice of parameters (shown in Figure~\ref{fig:bh_mass_rmp}a) predicts two strong peaks around 8 and 10 \msun, a tail of BHs above 18 \msun, and a significant dearth of BHs between 10 and 18 \msun.

We use the flexibility of the \citet{Maltsev+2025:2025AA...700A..20M} models to explore the impact of how fallback is implemented (panels (b) and (c) of Figure~\ref{fig:bh_mass_rmp}). Increasing the partial fallback fraction leads to a larger fraction of the ejecta falling back into the BH. This larger fraction increases the mass of BHs that formed via partial fallback, which fills in some of the underdensity between 10 and 18 \msun, at the expense of removing the subpopulation of BHs below 7 \msun (shown in Figure~\ref{fig:bh_mass_rmp}b), which are now shifted to higher masses. 
Furthermore, increasing the \textit{probability} of partial fallback has no effect on the regime between 10 and 18 \msun, but can significantly increase the rate of lower mass BHs (shown in Figure~\ref{fig:bh_mass_rmp}c). This increase is a result of lower mass stars forming BHs (instead of NSs) via partial fallback. However, the overall number of BHs is still 60\% of that predicted by the fiducial model (see Table~\ref{tab:summary_table}).

\subsubsection{Impact of varying the remnant mass prescription on BH kinematics}

The variations in remnant mass prescription not only change BH masses, as seen in Figure~\ref{fig:bh_mass_rmp}, but also vary the fraction of ejecta that falls back onto the collapsing core during a CCSN. A larger fallback fraction can negate natal kicks, as discussed in detail in Section~\ref{sec:fid_rates_binarity} and shown in Figure~\ref{fig:kicks-vs-mass}. As such, one should expect that variations in remnant mass prescriptions could impact BH kinematics, where larger natal kicks lead to thicker discs. We show the variations in the effective scale height of BHs as a function of remnant mass prescriptions in left panel of Figure~\ref{fig:summary-stats}, and now comment briefly on specific differences among the models.

\paragraph{\citet{Fryer+2012:2012ApJ...749...91F} \textit{rapid}} This variation produces a slightly smaller scale height than the fiducial model. This decrease is because the variation produces preferentially higher mass BHs (Figure~\ref{fig:bh_mass_rmp}a), which experience higher fallback, decreasing their kicks on average. 

\paragraph{\citet{Maltsev+2025:2025AA...700A..20M}} Each of these variations also predict lower scale heights than the fiducial model. This decrease stems not only from the higher average BH mass, but also because BHs are only formed with either complete fallback or partial fallback in this model. In this way, all natal kicks are at least partially damped. For the same reason, decreasing the partial fallback mass fraction tends to increase the BH scale height. The variation in which we set $P({\rm pf}) = 1.0$, which by construction has a much greater fraction of systems with partial fallback, has a larger scale height, consistent with the fiducial model.

\paragraph{\citet{Mandel+2020:2020MNRAS.499.3214M}}This variation is slightly more complicated since it changes both the remnant mass prescription and natal kick model. This variation predicts the smallest scale height of all those that we consider, at $368\unit{pc}$, nearing the scale height of stars drawn from our star formation history ($306\unit{pc}$). The small scale height is a combination of two effects: (1) natal kicks are generally weaker under this model \citep[see Figure 4 of ][]{Mandel+2020:2020MNRAS.499.3214M} so BHs kinematics are only weakly perturbed from their values at formation, and (2) the limit for BHs to receive full fallback is lower ($8\msun$ instead of $14\msun$ in \citet{Fryer+2012:2012ApJ...749...91F}), meaning that more BHs have their kicks nullified.

\subsection{Assuming single star evolution overestimates the scale height of BHs}\label{sec:binary-vs-single}

Earlier work predicting the positions of compact objects in the Milky Way has assumed that they all form as single stars \citepalias{underworld}. However, binary interactions could have an impact on both the number and scale height of BHs.
The total number may be altered due to mass transfer and stellar mergers changing which stars can form BHs. In terms of scale height, for single star models the only source of kinematic heating is the natal kick. However, the landscape is more complex when binary companions are accounted for: first, the natal kick applied to a BH in the fiducial model is dependent on its mass (see Figure~\ref{fig:kicks-vs-mass}), which can be altered via binary mass transfer or mergers. Second, the impact of the kick is different if the binary system remains bound. Third, BHs may also be ejected from binaries by their companion's SN, though these velocities are typically small compared to natal kicks (see Section~\ref{sec:fid_spatial}).

We quantify how binary interactions affect the results using a ``single star evolution'' variation on the fiducial model. We take the exact same initial population as the fiducial model, but evolve each companion as a separate, single star. This choice forces the initial stellar properties of each star to be identical to those in the fiducial population, but removes any impacts of binary interaction. 

Under the assumption of single star evolution, we find that the total number of Galactic BHs decreases by around 5\% relative to the fiducial model. Given that the frequency of stellar mergers (which can produce additional BHs from two low-mass stars) is high for typical binary models, one might have expected that the binary population model should have produced many more BHs than the ``single star evolution'' model. However, binary mergers can also \textit{reduce} the population of BHs, by merging two stars that each could have produced their own BHs, leaving only a single BH. We find that these two scenarios nearly balance each other, resulting in only a modest decrease in BH occurrence rate from single stars.


Despite the relatively small change in number of BHs when changing from a binary to single evolution models, the scale height of BHs increases. The scale height when assuming single star evolution increases by around 30\% to $1010 \unit{pc}$ from $790\unit{pc}$ in the fiducial model (see Figure~\ref{fig:summary-stats}). This increase is driven primarily by additional companion mass present in systems that remain bound after a core-collapse event. For a single star, the natal kick from core collapse is applied only to the resulting BH. However, if the binary remains bound after the core-collapse event, both the BH and (potentially massive) companion are impacted. Given the additional mass and identical total momentum imparted by the kick, the system velocity can decrease significantly \citep[e.g.,][Appendix A]{Pfahl+2002}. A secondary effect is that BHs are slightly more massive on average when accounting for binary evolution (due to mergers) and, as such, they are more likely to experience higher levels of fallbacks and weaker natal kicks.

\subsection{Accounting for the time-evolution of the Milky Way gravitational potential}\label{sec:time-evolving-potential}

The mass of the Milky Way has grown over time and as such, early in the history of the Galaxy, the gravitational potential was weaker. BHs have extremely long lifetimes, are often formed early in the history of the Galaxy, and may receive natal kicks. For these reasons, BH kinematics could be sensitive to long-term evolution of the Galactic potential.

In order to assess the effects of the evolution of the potential, we construct simple a time-evolving model for the Galactic potential (see Appendix~\ref{app:tep-model}). We recompute the Galactic orbits of BHs formed in the fiducial model in this time-evolving potential.

The scale height of BHs increases by around 20\% when we account for the mass growth of the galaxy over time. The weaker potential at the time of earlier kicks allows the BHs to reach larger distances from the Galactic plane. One may expect that, since the mass of the galaxy has increased by a factor of $f = 2.5$ in our model, the difference between the time-evolving case and static potential should be larger. However, the adiabatic increase in Galaxy mass damps orbits, such that the expected increase scales approximately as $f^{1/4}$, as we detail in Appendix~\ref{app:analytic-tep}. In this way, BHs may initially reach large heights, but by present day their orbits are damped to be closer to the Galactic plane.

The fraction of BHs that escape the Galaxy increases by $2.5 \times$, from 2.7\% to 6.9\% when accounting for the time evolution of the galaxy. The increase in escape fraction is driven by the lower mass, and therefore escape velocity, of the Galaxy at early times. The lower threshold allows a weaker kick to cause a BH to escape. Although the increase in free-floating BHs is difficult to observe, this effect may also increase the escape fraction and offset distribution of binary neutron stars and the short gamma-ray bursts they may produce upon merger \citep[e.g.,][]{Zevin+2020:2020ApJ...904..190Z}.

\subsection{Milky Way BHs are mostly robust to uncertainties in binary mass transfer}\label{sec:mt-variations}

The stability of mass transfer, efficiency of stable mass transfer, and efficiency of common-envelope phases remain critical uncertainties in binary physics \citep[e.g.,][]{Soberman+1997,Ivanova+2013, Gallegos-Garcia+2021:2021ApJ...922..110G,Ropke+2023:2023LRCA....9....2R,Marchant+2024:2024ARA&A..62...21M,Lechien+2025:2025ApJ...990L..51L,Sen+2026:2026ApJ..1000....2S}.
However, we find that these uncertainties do not have a strong impact on our results (see Appendix~\ref{app:mt_variations}). We find that the scale heights vary by no more than 5\% and the correlations between BH mass and Galactic height remain present. This indifference is likely because BH kinematics are relatively insensitive to orbital velocities (see Section~\ref{sec:kick-variations}), which are varied by mass transfer episodes. The total number and mass of BHs in these variations is more sensitive, varying by no more than $20\%$. The largest difference comes from assuming that mass transfer is always nonconservative, thus preventing the formation of many BHs that would form after the progenitor accretes from its companion.

\section{Black holes in binaries}\label{sec:bh_in_binaries}

In the previous sections we have considered the distributions of the overall Milky Way BH population, which is dominated by isolated BHs that are primarily detectable via microlensing. We now turn our focus to the small fraction of BHs which instead remain in bound binary systems at present day. Despite the intrinsic paucity of BHs in binaries, they have several more avenues for potential detection. BHs in binaries can be identified by their interaction with a visible partner, whether by the accretion of material \citep[as an X-ray binary, e.g.,][]{Webster+1972:1972Natur.235...37W, Bolton+1972:1972Natur.235..271B} or influence on the motion of the companion \citep[detectable via astrometry or radial velocities, e.g.,][]{El-Badry+2023:2023MNRAS.518.1057E_GaiaBH1}. Therefore, BHs in binaries are a potentially more observable subpopulation, despite their rarity.

We predict the occurrence rates of each binary type by scaling our results to a full Milky Way star-forming mass of $\num{6e10}\msun$ \citep{Licquia+2015}.
In the fiducial model, the most common BH binary companion is another BH ($\num{6.7e6}$), followed by a white dwarf (WD, $\num{1.3e6}$), and in rare cases a NS ($\num{1.4e5}$), or a luminous stellar companion ($\num{1.7e5}$).

These relative rates are driven by a combination of formation rates, natal kicks, and lifetimes. BHs in binary systems are preferentially higher mass (see Section~\ref{sec:bh-binary-mass-skew}) and, given that we assume the mass ratio distribution is uniform, progenitors of BHs in binaries are more commonly paired with massive stars, which also form BHs. As a result, BH-BH binaries are the most common form of BH binary. A more unequal mass ratio system (where the secondary could produce a NS or WD) is more likely to experience an unstable common-envelope event which often results in mergers, decreasing the rate of these systems. The next most common initial pairing is BH and NS progenitors, however NSs often have strong natal kicks, which unbind most systems. As a result, BH-WD systems are around $10\times$ more common than BH-NSs. BH-star systems are similarly rare. These systems face several difficulties: pairings with massive stars are short-lived due to the stars' short lifetimes, whilst those with low-mass stars must either survive a common-envelope event, or remain wide enough to avoid mass transfer (while also potentially surviving a supernova natal kick).

\subsection{The occurrence rates of BH binaries are most sensitive to natal kicks and initial distributions}

\begin{figure*}
    \centering
    \includegraphics[width=\textwidth]{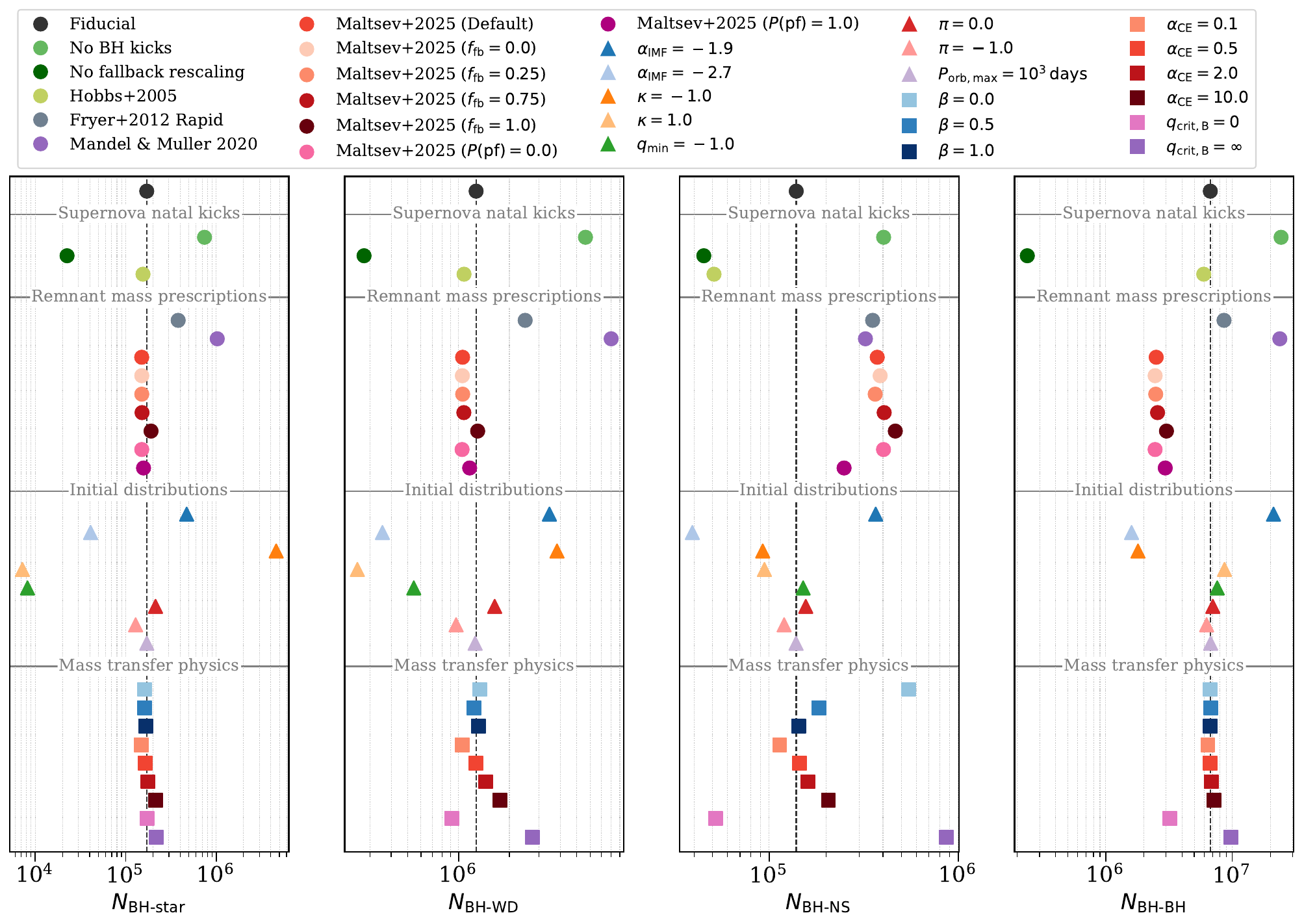}
    \caption{The number of Milky Way BH binaries for our suite of simulations, described in Section~\ref{sec:variations}. Each panels show the total number of BH-star, BH-white dwarf, BH-NS, and BH-BH binaries respectively. Variations of natal kicks and remnant mass prescriptions are shown as circles, variations of initial distributions are triangles, and variations of mass transfer physics are circles.}
    \label{fig:bh-binary-rates}
\end{figure*}

The total number of Milky Way BH binaries at present day are sensitive to the assumptions we make in our simulations. In Figure~\ref{fig:bh-binary-rates}, we show how the total number of BH-star, BH-WD, BH-NS, and BH-BH binaries at present day in the Milky Way varies for our suite of variations. The values are tabulated in Table~\ref{tab:bh_binaries}. 

The most impactful variations are those that change the initial distributions of binaries. In particular, changing the initial mass ratio distribution to be skewed towards equal mass ratio systems disfavours the formation of all BH binaries except BH-BH systems. This variation produces the lowest number of BH-WD and BH-star systems, with the latter decreasing by more $50\times$. The decrease happens because BH progenitors are much less frequently paired with low-mass companions. Limiting the minimum mass ratio based on the pre-main sequence lifetime of the companion has a similar, but lesser, effect on each binary type. As we noted in Section~\ref{sec:variations}, altering the slope of the IMF has a strong effect on the rate of BH formation, which also translates to the rate of BH binary formation. A top-heavy IMF ($\alpha_{\rm IMF} = -1.9$, \citealt{Schneider+2018:2018Sci...359...69S}) produces more BHs and proportionally more of them pair with each type of companion, whilst a bottom-heavy IMF produces some of the lowest totals for each binary type.

The number of BH binaries is also very sensitive to the treatment of BH natal kicks. Removing natal kicks for all BHs increases the number of most binary types by a factor of ${\sim}4$ (BH-NS are less strongly affected since NSs still have strong kicks). In this variation, disruptions are only possible due to the kick of a NS companion, or the Blaauw kick from symmetric mass loss. Conversely, giving BHs the same kicks as NSs drastically reduces the numbers of all binaries, as one may expect from kicks causing disruptions. In particular, this variation forms $30\times$ fewer BH-BH binaries, the lowest total number for all variations we consider. We discuss how this could be probed with microlensing in Section~\ref{sec:discussion}.

Most changes to the remnant mass prescription only slightly affect the number of BH binaries. The main effect is that the variations on the fiducial model tend to produce more NSs and fewer BHs (increasing the number of BH-NSs and decreasing the number of other systems). However, it is notable that the \citet{Mandel+2020:2020MNRAS.499.3214M} prescription produces one of the highest totals of BH-star, BH-WD, and BH-BH binaries. BH natal kicks are drawn from a distribution that is skewed to much weaker kicks in this model. Additionally, the threshold for complete fallback is much lower (a carbon-oxygen core mass of $8\msun$, compared to $14\msun$ in the fiducial \citet{Fryer+2012:2012ApJ...749...91F} delayed model, see also Figure~\ref{fig:bh-binary-masses}), meaning that a much larger fraction of kicks are nullified. Overall, this means that the natal kicks applied to BHs are generally much lower, allowing them to remain in bound systems much more frequently. It is therefore possible that observations of BH-star binaries may offer strong constraints on the viability of this prescription, though this requires a more detailed exploration of which BH-star binaries are observable by astrometric or radial velocity searches.

The total number of BH binaries is mostly robust to the variations of mass transfer physics. The exception is the BH-NS population, which is particularly sensitive to the stability of mass transfer and the efficiency of stable mass transfer. These systems are almost never formed without an episode of mass transfer (around 1\%), whereas BH-BHs and BH-WDs are formed without any Roche-lobe overflow around 60\% of the time. Forcing mass transfer to always be unstable causes more common-envelope events, resulting in more mergers (see $f_{\rm merger}$ column in Table~\ref{tab:mt_summary_table} and Figure~\ref{fig:summary-other-cols}) instead of BH-NS formation. For a similar reason, increasing common-envelope efficiency, which makes it easier to survive a common envelope and avoid merger, produces more BH-NSs. Decreasing the efficiency of mass transfer increases the rate of BH-NS formation since it prevents secondary stars from accreting material, limiting their expansion to prevent reverse case B mass transfer. This limited expansion means that mass transfer occurs as case C and results in a successful common envelope rather than a merger.

\subsection{BHs in binaries are skewed to higher masses}\label{sec:bh-binary-mass-skew}

\begin{figure}
    \centering
    \includegraphics[width=\columnwidth]{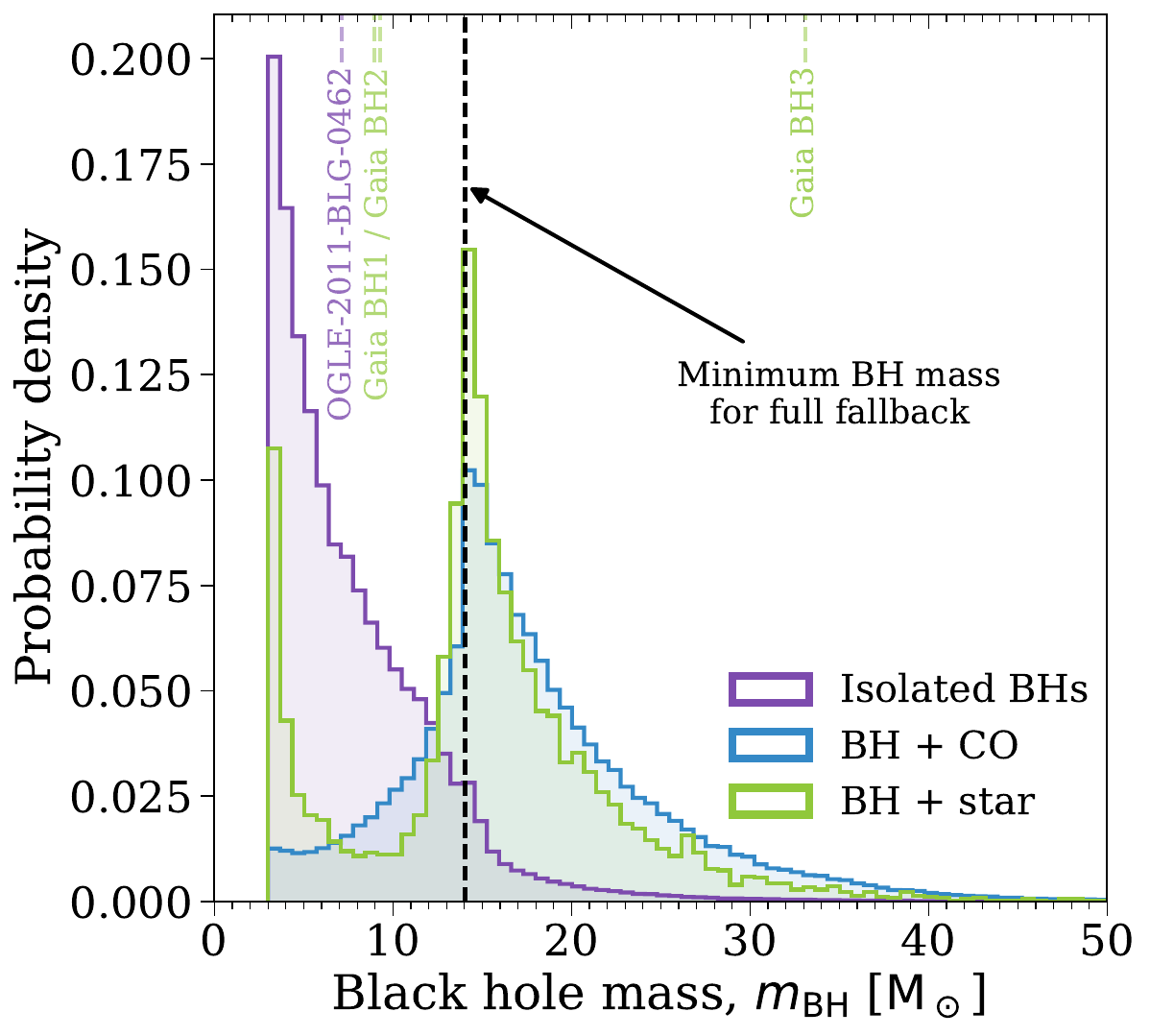}
    \caption{BHs in binaries are heavily skewed to higher masses. The distribution of BH masses in the Milky Way, separated into isolated BHs (purple), those in a binary system with a luminous companion (green), and those in a binary system with another compact object (blue); these three classes roughly correspond to the sources that would potentially be detectable with different techniques (microlensing, astrometry and/or XRBs, and gravitational waves, respectively). The black dashed line shows the minimum mass that results from full fallback after a core-collapse event (resulting in no natal kick, which could disrupt a binary). We annotated the values of the isolated BH OGLE-2011-BLG-0462 and \gaia BHs with dashed lines, coloured corresponding to the histogram to which they would belong.}
    \label{fig:bh-binary-masses}
\end{figure}

In Section~\ref{sec:mass-from-rmp} we highlighted how the Milky Way BH mass distribution may be a helpful calibrator for remnant mass prescriptions. However, the intrinsic mass distribution of BHs differs for isolated and bound systems, as we highlighted in Figure~\ref{fig:masses-by-type}. In Figure~\ref{fig:bh-binary-masses}, we show how the mass distribution varies for different subsets of BHs. The isolated BH mass distribution has the same shape as the total distribution we showed in Figure~\ref{fig:bh_mass_rmp}a (because the majority of BHs are isolated) with a median mass of 6.6\msun.

In contrast, BHs in bound binaries are biased towards higher masses, which experience weaker natal kicks (see Figure~\ref{fig:kicks-vs-mass}). BHs with a stellar companion (green distribution) have a dearth of masses below 10 \msun and instead the distribution peaks around 14\msun, with an average BH mass of 15\msun. This peak occurs because we assume that BH natal kicks are rescaled by fallback, which increases at higher masses. For the \citet{Fryer+2012:2012ApJ...749...91F} \textit{delayed} prescription that we use in the fiducial model, the minimum BH mass for full fallback is around 14 \msun (see Figure~\ref{fig:kicks-vs-mass}), which we show as a dashed line in Figure~\ref{fig:bh-binary-masses}. BHs above this mass receive no natal kick and are therefore more likely to remain in bound binaries. For the same reason, BHs in binaries with another compact object (blue distribution), which may need to survive \textit{two} natal kicks, are further skewed to higher masses, with an average mass of 16.4\msun. Note that these BH-BH and BH-NS binaries do not correspond to those detected by gravitational waves, since we have no condition that they are \textit{merging} binaries (and are not applying selection effects, which are sensitive to mass).

We additionally show the confirmed population of BHs observed by \gaia and detected by microlensing with annotated dashed lines along the top axis. OGLE-2011-BLG-0462, the BH detected via microlensing, has a mass of $7.15\pm0.83\msun$ \citep{Sahu+2025:2025ApJ...983..104S}, which is typical for our simulated distribution of isolated BHs. The masses of the \gaia BHs are in rare areas of parameter space in our simulated model, however this is not unexpected given the \gaia selection criteria may bias observations to certain values \citep[e.g.,][]{El-Badry+2024:2024OJAp....7E.100E}.

\subsection{BH mass in BH-star binaries is tightly correlated with orbital period and Galactic orbits}

\begin{figure}
    \centering
    \includegraphics[width=\columnwidth]{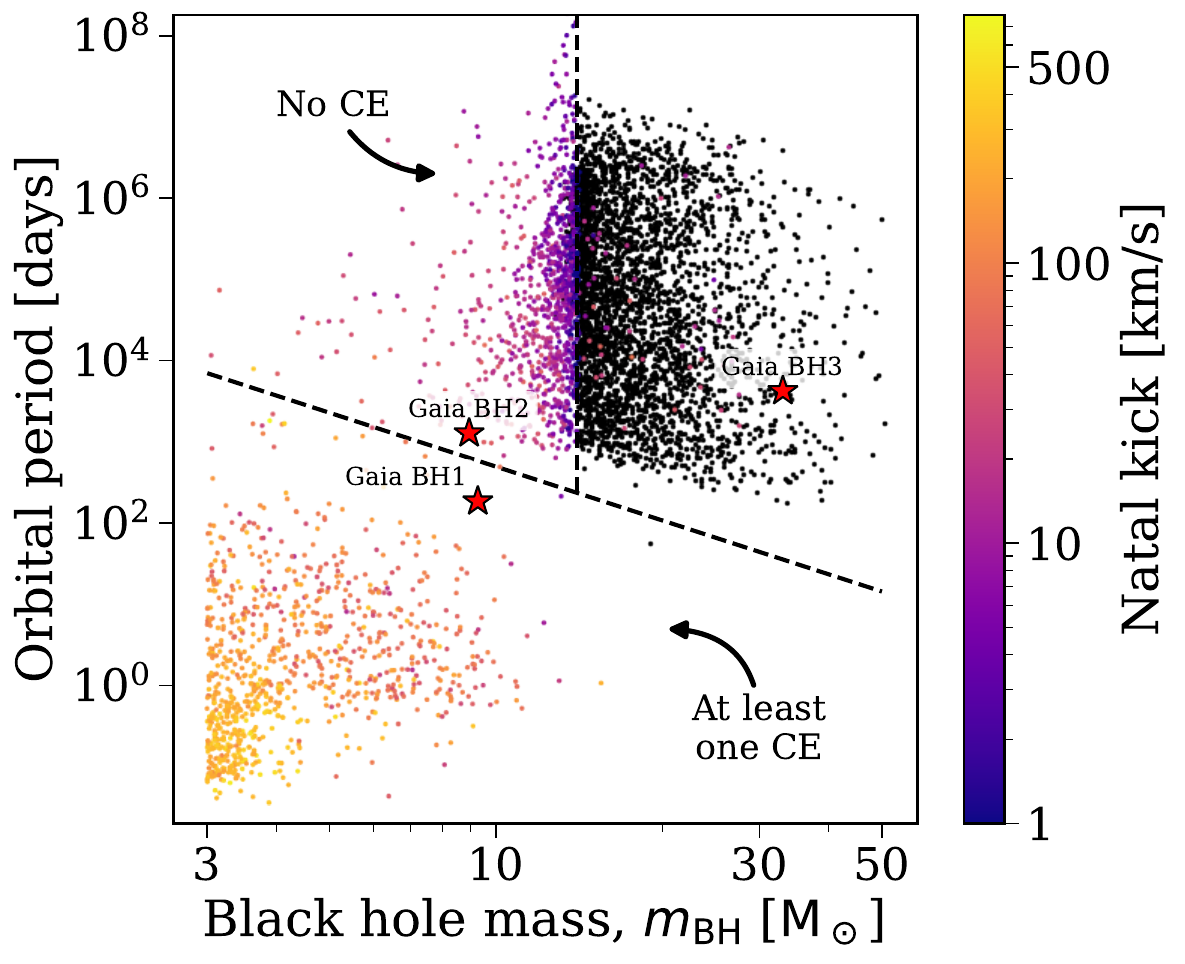}
    \caption{BHs with luminous companions have a bimodal distribution in mass and orbital period. Low-mass BHs are found in close binaries (after a common-envelope event) and receive strong natal kicks, while high-mass BHs receive little to no kick and remain in wide, non-interacting binaries. Black points are BHs that achieved full fallback and therefore received no natal kick. The diagonal dashed line separates systems that have experienced a common envelope (below) from those that have had no mass transfer (above). We add points for the 3 \gaia BH detections.}
    \label{fig:bh-porb-kick-correlations}
\end{figure}

The bimodality in BH mass for BH-star binaries (see Figure~\ref{fig:bh-binary-masses}) is also reflected in the orbital period of these systems \citep[e.g.][]{Breivik+2017:2017ApJ...850L..13B, Chawla+2022:2022ApJ...931..107C}. In Figure~\ref{fig:bh-porb-kick-correlations}, we show how BH mass correlates with orbital period. We colour each point by the natal kick magnitude from the core-collapse event that produced the BH in the system, where black points indicate that the BH received (effectively) no natal kick\footnote{We do not account for the natal kick from asymmetric neutrino emission. These kicks are not expected to exceed $10\unit{km}{s^{-1}}$ \citep[e.g.][]{Vigna-Gomez+2024:2024PhRvL.132s1403V} and as such would be unlikely to unbind these binaries.} (i.e., full fallback). We find that low-mass BHs, which preferentially receive strong natal kicks (see Figure~\ref{fig:kicks-vs-mass}), are found in close binaries with orbital periods below 100 days \citep{Chawla+2022:2022ApJ...931..107C}. These systems underwent at least one common-envelope event, which shrinks the orbit and increases the binding energy of the system, making it able to survive a strong natal kick. In contrast, high-mass BHs, which receive little to no natal kick, are found in wide binaries with orbital periods of typically at least 1000 days, with a large spread up to $10^8$ days. These wide binaries are non-interacting, in that there was no mass transfer episode. We note that these binaries are still not so wide that Galactic tides could strongly perturb their orbit \citep[see Figure 2 of][]{Stegmann+2024:2024arXiv240502912S}. Close binaries are disfavoured for high-mass BHs because their progenitors would expand enough to engage in (unstable) mass transfer with their companion, leading to a merger.


The separation between the BH-star systems that experienced a common envelope from those that had no mass transfer\footnote{Very few BH-star systems that we simulated experienced only stable mass transfer.} is quite clearly separated in the fiducial model. The present-day orbital period threshold for whether the system underwent a common envelope, shown as a diagonal dashed line in Figure~\ref{fig:bh-porb-kick-correlations}, can be written as
\begin{equation}
    \frac{P_{\rm orb}}{7000 \unit{days}} < \qty(\frac{m_{\rm BH}}{3 \msun})^{2.2} \implies \text{Had mass transfer}
\end{equation}
This condition accurately predicts the mass transfer history 99.7\% of the time for the fiducial model. However, we caution that this orbital period threshold may be altered by updated stellar radii prescriptions (which affects the occurrence of mass transfer) or variations in mass transfer physics (which affects the post-mass-transfer orbital period).

The detected \gaia BHs occupy areas of parameter space that are poorly populated by the intrinsic population. In Figure~\ref{fig:bh-porb-kick-correlations}, we add a red star for each the observed \gaia BH systems. In particular, \gaia BH1 and BH2 are both found close to the region in $\{m_{\rm BH}, P_{\rm orb}\}$ that separates the populations that have experienced mass transfer or not. Our simulated population is therefore mildly in tension with this data, but we note that \gaia selection effects disfavour very short and long periods. Ultimately, we require the larger dataset of BH-stars that we expect \gaia DR4 will deliver in order to understand the severity of this tension.

\begin{figure}
    \centering
    \includegraphics[width=\columnwidth]{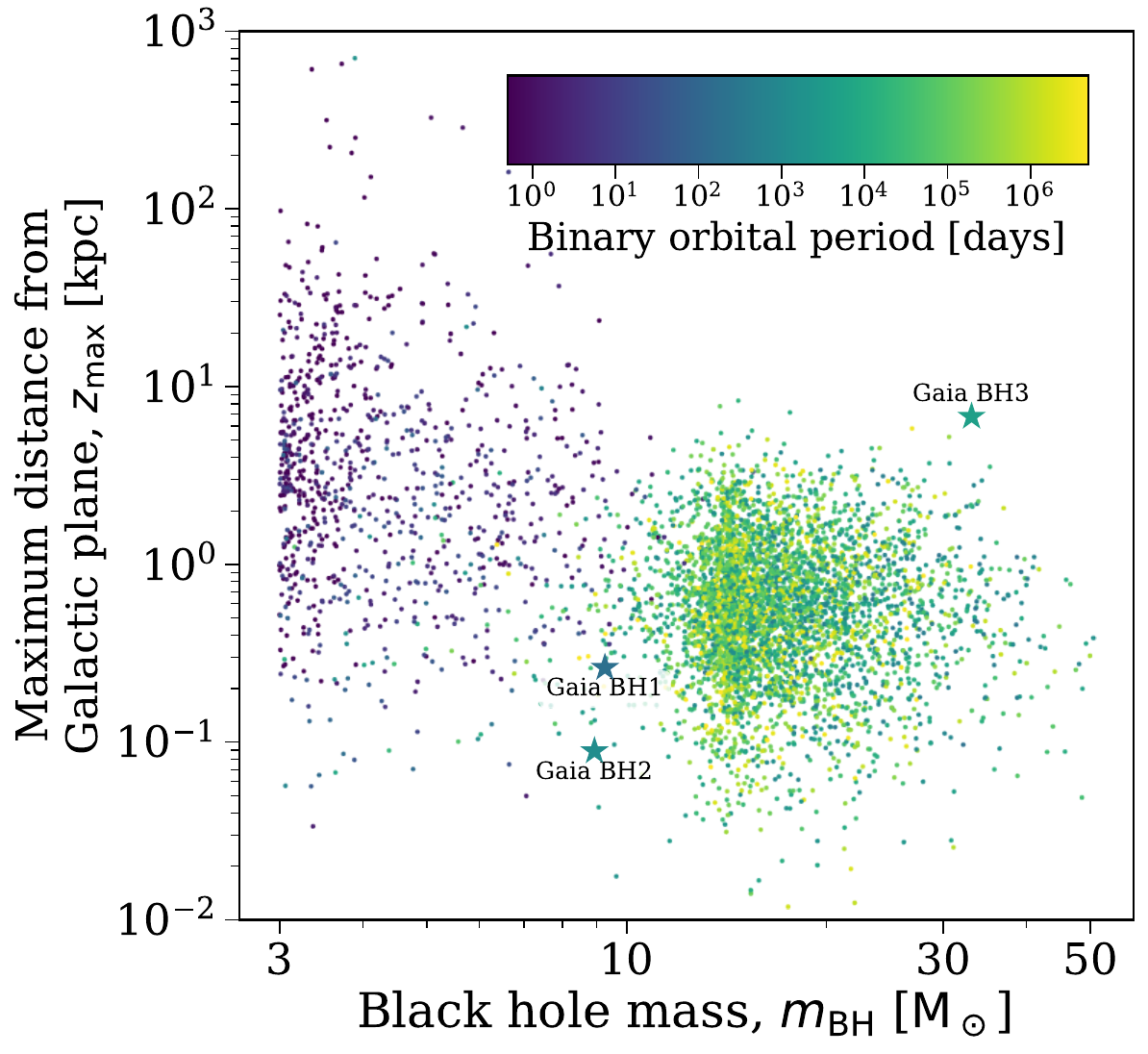}
    \caption{The spatial distribution of BH-star binaries is strongly influenced by the mass of the BH. Points show the distribution of BH mass and the maximum distance from the Galactic plane achieved on the binary's orbit around the Galaxy, coloured by binary orbital period. We also add points for the 3 \gaia BH detections.}
    \label{fig:bh-star-spatial}
\end{figure}

Another implication of the correlation between orbital period and BH mass is that the Galactic orbits of tight, low-mass BH-star binaries is significantly more heated than for wide BH-star binaries. In Figure~\ref{fig:bh-star-spatial}, we show how the mass of BHs in BH-star binaries correlates with their maximum height above the Galactic plane during their Galactic orbit. We find that tight (more blue) BH-star binaries reach significantly larger heights than wide (more green/yellow) BH-star binaries.

We additionally add star markers for the detected \gaia BH systems. We integrated their Galactic orbits forwards from the orbital elements reported in each discovery paper \citep{El-Badry+2023:2023MNRAS.518.1057E_GaiaBH1, El-Badry+2023:2023MNRAS.521.4323E_GaiaBH2, GaiaCollaboration+2024:2024AA...686L...2G_GaiaBH3} through the same Galactic potential as the fiducial model for $5 \unit{Gyr}$, and recorded the maximum distance from the plane for each. \gaia BH1 and BH2 have a $z_{\rm max}$ that is not atypical for their mass range, and have regular disc-like Galactic orbits. In contrast, \gaia BH3 is an outlier in terms of $z_{\rm max}$ compared to our intrinsic fiducial population. This deviation is expected as BH3 is thought to have formed in a globular cluster rather than the Milky Way disc \citep[e.g.,][]{Balbinot+2024:2024A&A...687L...3B, MarinPina+2026:2026arXiv260424874M}. Indeed our results are supportive of this scenario, since we find that no disc-origin BH of BH3's mass can attain such a large $z_{\rm max}$.

Finally, we estimate the number density of BH-star binaries in the solar neighbourhood with a torus with a major radius of $8\unit{kpc}$ and a minor radius of $0.5 \unit{kpc}$. We calculate this density as $\num{2e-7} \unit{pc^{-3}}$, which implies that the nearest BH-star binary to the Sun is, on average, ${\sim}170 \unit{pc}$ away. For comparison, \gaia BH-1, the closest detected BH-star binary, is located around $480\unit{pc}$ from Earth.

\subsection{BHs undergoing accretion at present day}

We find that approximately 7\% of BHs in BH-star systems are undergoing accretion via Roche-lobe overflow from their stellar companion at present day. This fraction corresponds to a total of around 11,000 systems in the Milky Way. From Figure~\ref{fig:bh-porb-kick-correlations}, one can see that selecting close binaries (necessary for mass transfer) biases the population to low-mass BHs. As a result of this bias, the average BH mass for BHs undergoing accretion is $3.8\msun$, though the distribution extends up to $25 \msun$. Companions are also typically low-mass stars (which are longer-lived and thus more likely to be fusing at present day), with an average mass of $0.6\msun$. The number of these systems generally scales with the number of BH-star systems. For example, if BHs always receive the same kicks as NSs (i.e.\ no fallback rescaling), we predict there would only be around 1500 BHs accreting from a stellar companion in the entire Milky Way. Thus XRB detection rates could offer constraints on uncertain binary physics (see left panel of Figure~\ref{fig:bh-binary-rates}).

However, we caution that these accreting BHs are not necessarily detectable XRBs. Converting the intrinsic population to the detected XRBs would require a more detailed treatment of the expected X-ray luminosity \citep[e.g.,][]{Misra+2023}. Accounting for selection effects may skew the distribution of masses compared to the results we report above. An upcoming study will examine the demographics and kinematics of detectable Milky Way XRBs in more detail \citep{davids_xrbs}.

\section{Implications for the observational BH landscape}\label{sec:discussion}

We have presented predictions for the intrinsic population of Milky Way BHs in this work. A full consideration of the \textit{observable} population of these BHs will offer exciting insights from upcoming BH datasets.

\paragraph{Microlensing with Roman}

The \romanFull is expected to detect 100s of isolated BHs \citep[e.g.,][]{Lam+2022:2022ApJ...933L..23L} via microlensing. These isolated BHs represent the vast majority of the population (see Section~\ref{sec:fid_results}) and thus our statements regarding the constraints offered from BH positions and masses hold for these systems. With 100s of BH mass measurements, we expect that \romanShort may be able to offer insights into remnant mass prescriptions and rule out some of the distributions shown in Figures~\ref{fig:bh_mass_rmp} and \ref{fig:rmp-cdfs}. In particular, \romanShort could test the strong predictions from the variation using the \citet{Maltsev+2025:2025AA...700A..20M} remnant mass prescription, which suggests almost no BHs will be detected with masses between $10$ and $18\msun$ (see Section~\ref{sec:rmp_effect_on_bh_mass}).
Given the constrained direction of the \romanShort bulge time domain survey, it may be difficult to assess how the BH mass correlates with distance from the Galactic plane (see Figure~\ref{fig:average-mass-by-z}). However, it's possible that comparison of the \romanShort population of BHs to any detected at larger distances from the plane (such as microlensing surveys focused on the LMC/SMC or all-sky photometric surveys) could help to constrain BH natal kicks.

\romanShort may also be sensitive to BH-BH systems as binary lenses. The detectability of these systems \textit{as} binary lenses is very sensitive to the angular separation of the binary, $\theta_{\rm sep}$ \citep[e.g.,][]{Schneider+1986:1986A&A...164..237S, Gaudi+2012:2012ARA&A..50..411G}. Specifically, systems with $\theta_{\rm sep} \approx \theta_E$, where $\theta_E$ is the Einstein radius are most detectable as binaries. Close systems with $\theta_{\rm sep} \ll \theta_E$ tend to be disguised as a single, more-massive lens, while wide systems with $\theta_{\rm sep} \gg \theta_E$ often only have one component producing a lensing effect on a source. Whether an individual lens system has strong deviations from a single lens model is dependent both on its configuration and on the source trajectory. In particular, lenses with $0.7 < \theta_{\rm sep} / \theta_E < 1.8$ have a 50\% chance of being detected as a binary lens \citep{Han+1999:1999MNRAS.308.1077H, Han+2001:2001MNRAS.325.1281H}. We estimate the fraction of our BH-BH population that fulfils these criteria by calculating their Einstein radii and angular separations. We select the subset of BH-BHs with Galactocentric radii $R < 8.2 \unit{kpc}$, in a cone such that $|z| < (8.2 \unit{kpc} - R) \tan(5^\circ)$. We approximate their Einstein radii as
\begin{equation}\label{eq:approx_einstein}
    \tilde{\theta}_E \sim \sqrt{\frac{4 G M}{c^2}} \qty(\frac{1}{d} - \frac{1}{8.2 \unit{kpc}}),
\end{equation}
where $M$ is the total mass of each binary, $d$ is its distance from a circle with $(R, z) = (8.2, 0)\unit{kpc}$, and we have assumed every source is located at $8.2 \unit{kpc}$.

\begin{figure}
    \centering
    \includegraphics[width=\columnwidth]{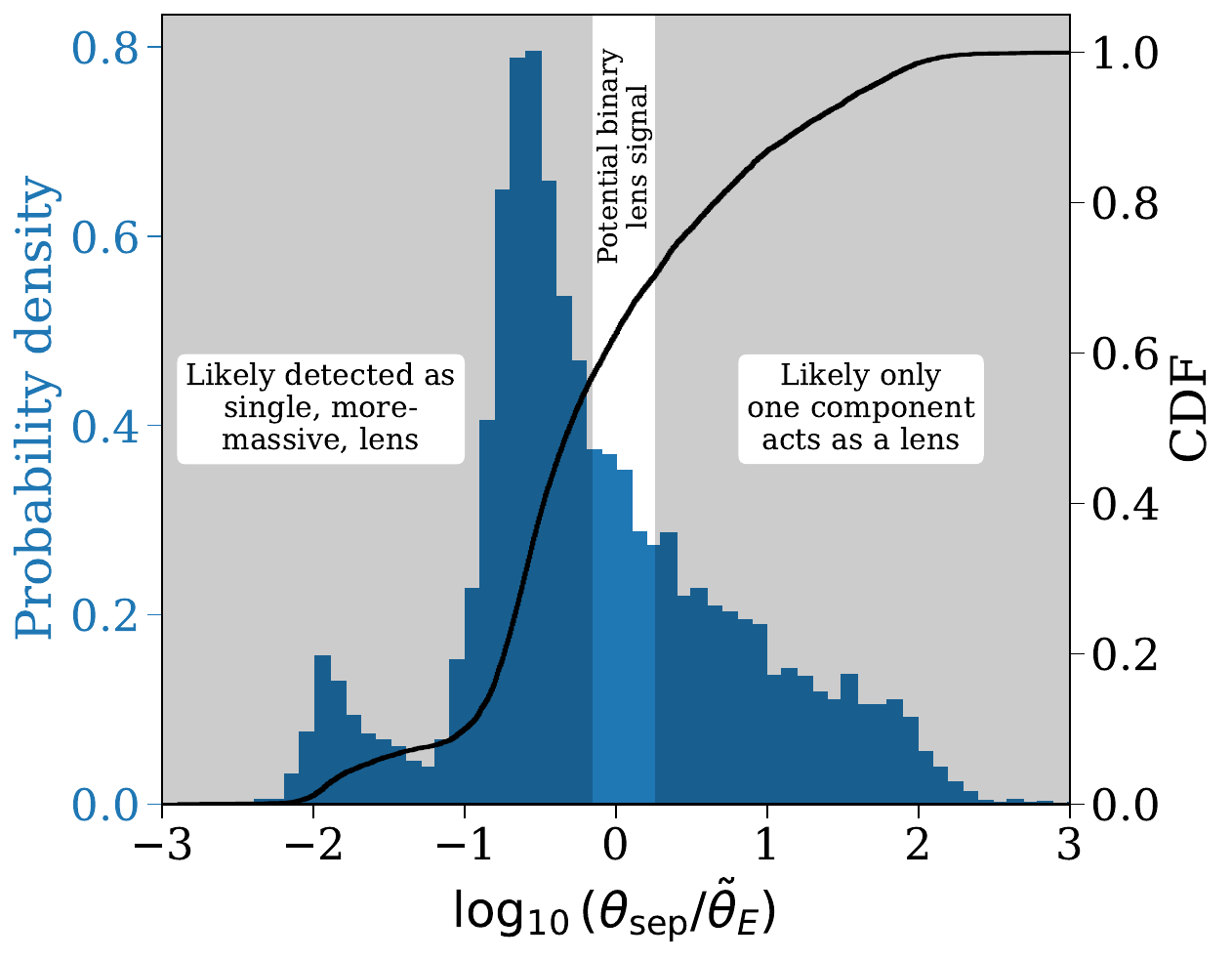}
    \caption{A fraction of BH-BH binaries may be detectable as binary lenses with microlensing. The blue histogram shows the distribution of the ratio between the angular separation, $\theta_{\rm sep}$, and approximate Einstein radius, $\tilde{\theta}_E$, (see Eq.~\ref{eq:approx_einstein}) for simulated BH-BHs. We overlay the normalised cumulative distribution function of this population as a black line (using the right y-axis). The annotations indicate the regions in which a BH-BH lens likely appears as a binary lens or not. The central unshaded region indicates the range in which there is a 50\% chance of detecting a binary lens signal.}
    \label{fig:bh-binary-lenses}
\end{figure}

We show the distribution of $\theta_{\rm sep} / \tilde{\theta}_E$ for the subset of BH-BHs in the cone towards the Galactic centre in Figure~\ref{fig:bh-binary-lenses}. We find that around $13\%$ of these sources have $0.7 < \theta_{\rm sep} / \theta_E < 1.8$ (shown as the unshaded region), and thus a 50\% chance of being detected as a binary lens. We caution that this distribution does not take into account that some systems are more likely to serve as lenses than others as a function of their separation, mass, and distance \citep[e.g.][]{Abrams+2025:2025ApJ...980..103A}. Though, this fraction suggests that a more detailed investigation (accounting for the distributions of sources and selection effects) into \romanShort's ability to characterise the population of BH-BH binaries is worthwhile \citep[e.g.,][]{Lam+2020:2020ApJ...889...31L, Abrams+2025:2025ApJ...980..103A}. This may be particularly useful because their formation rates are highly sensitive to natal kicks and may offer constraints on this uncertain physics (see Figure~\ref{fig:bh-binary-rates}).

Overall, it is necessary to connect our results with tools such as \texttt{SPISEA} \citep{Hosek+2020:2020AJ....160..143H}, \texttt{PopSyCLE} \citep{Lam+2020:2020ApJ...889...31L, Rose+2022:2022ApJ...941..116R, Abrams+2025:2025ApJ...980..103A}, and \texttt{SynthPop} \citep{Kluter+2025:2025AJ....169..317K, Huston+2026:2026arXiv260312219H}, in order to fully predict the detectable microlensing population for comparison with \textit{Roman} observations. This will be addressed in a series of papers on the BH (J.\ Lu et.\ al.\ in prep.) and binary (N.\ Abrams et.\ al.\ in prep.) populations.

\paragraph{BH-star binaries with Gaia DR4}

Three BHs in wide orbits have been discovered in \gaia observations to date \citep{El-Badry+2023:2023MNRAS.518.1057E_GaiaBH1, El-Badry+2023:2023MNRAS.521.4323E_GaiaBH2, Chakrabarti+2023:2023AJ....166....6C,  GaiaCollaboration+2024:2024AA...686L...2G_GaiaBH3}, with dozens more expected in \gaia DR4 later this year \citep[e.g.,][]{El-Badry+2023:2023MNRAS.518.1057E_GaiaBH1}. We predict that there are around 171,000 BH-star binaries in the Milky Way at present day \citep[similar to previous work, e.g.,][]{Breivik+2017:2017ApJ...850L..13B, Chawla+2022:2022ApJ...931..107C}. \gaia will not be sensitive to a large fraction of this population, for example because of a dim magnitude of the stellar companion, or the orbital period not matching the requirements for \gaia. Binaries with an orbital period greater than $10\unit{yr}$ will not complete an orbit during the \gaia mission, though a full orbit is not always required to classify a BH-star system \citep[e.g,][]{GaiaCollaboration+2024:2024AA...686L...2G_GaiaBH3}. These selection criteria will introduce biases to the population, particularly in terms of orbital period. The correlation between mass and orbital period therefore means that the BH mass distribution will be altered for the observable population (likely to lower mass BHs since long periods are disfavoured, see Figure~\ref{fig:bh-porb-kick-correlations}) and the systems will be typically found further from the Galactic plane (Figure~\ref{fig:bh-star-spatial}). For BHs in XRBs, which require even closer orbital periods, this skew to low BH masses will likely be even stronger. These biases should be taken into account when using these populations to constrain (binary) stellar evolution. A future study will present a full forecast for \gaia DR4 observations that combines of our simulations with tools such as \texttt{GaiaMock} \citep{El-Badry+2024:2024OJAp....7E.100E}.

\paragraph{Other BH binary discovery techniques} Radial velocity measurements have been used to confirm and constrain the measurements of the \gaia astrometric BHs \citep{El-Badry+2023:2023MNRAS.518.1057E_GaiaBH1, El-Badry+2023:2023MNRAS.521.4323E_GaiaBH2, Chakrabarti+2023:2023AJ....166....6C,  GaiaCollaboration+2024:2024AA...686L...2G_GaiaBH3}. The technique has also been used to discover a black hole candidate in the LMC \citep{Shenar+2019} and there are ongoing efforts to discover more \citep[e.g.][]{Yi:2019ApJ...886...97Y, Shenar:2024A&A...690A.289S, Lam:2026ApJ..1000..148L}. These efforts will be useful not only for confirming new \gaia BH candidates, but also supplementing discovery efforts. Additionally, one can use self-lensing to detect BH binaries. Self-lensing occurs in binary systems that are edge-on containing a compact object in which the compact object repeatedly lenses its luminous companion \citep{Leibovitz:1971A&A....15..251L, Maeder:1973A&A....26..215M}. There have been 5 self-lensing WD-main sequence binaries detected via Kepler all at $\sim 1$ AU \citep{Kruse:2014Sci...344..275K, Kawahara:2018AJ....155..144K, Masuda:2020IAUS..357..215M} and 1 TESS candidate \citep{Sorabella:2026ApJ...997....3S}, but currently no systems with a BH. There have been unsuccessful searches in ZTF \citep{Crossland:2024OJAp....7E..67C} and an ongoing citizen science project with SuperWASP data (A. McMaster et al.~in prep). However, there are predictions that 1--10s of these systems should be found in ZTF, TESS, and Rubin \citep[e.g.][]{Wiktorowicz:2021MNRAS.507..374W, Yamaguchi:2024PASP..136l4202Y, Nir:2025ApJ...978..169N,  Sajadian:2026AJ....171..330S}. Whether the searches are successful or not, they can put limits on the number of BH-star and BH-WD systems in the Galaxy.

\paragraph{Potential contamination from NSs} This study has focused on the population of BHs in the Milky Way, but NSs represent another kinematically heated population that is prevalent in the Galaxy. Our simulation data also encapsulate this population and, though we defer a full analysis to a later dedicated study, it is worth considering whether they could contaminate BH detections. We find that the Milky Way forms around $2.9\times$ more NSs than BHs, though due to their higher escape fraction there are only $2.6\times$ more NSs than BHs in the Milky Way at present day. These NSs are lower in mass and typically moving faster than BHs and, as such, their Einstein crossing times for microlensing events are typically much shorter than BHs. Therefore, despite their abundance, NSs should be separable from BHs in microlensing detections \citep[e.g.,][Figure 6]{Koshimoto+2024:2024ApJ...973....5K}.

\section{Conclusions}\label{sec:conclusions}

In this paper we have presented predictions for the population of Milky Way BHs, using self-consistent population synthesis and galactic dynamics simulations from \cogsworth. We have investigated the scale height, escape fraction, mass distribution, and binary properties of BHs. Moreover, we have considered a wide-range of variations beyond the fiducial model, exploring changes in remnant mass prescriptions, natal kick models, binarity, initial distributions, binary mass transfer physics, and the time-evolution of the Galactic potential. Our main findings can be summarised as follows:

\begin{conclusions}
    \conclusion{%
        The scale height of BHs is significantly larger than that of the visible galaxy
    }{
        In agreement with previous work, we find that BHs are more spatially extended than the visible galaxy by $2.5\times$, with a scale height of ${\sim}790\unit{pc}$ (see Section~\ref{sec:fid_spatial} and Figure~\ref{fig:scale-heights}).
    }

    \conclusion{%
        BHs located further from the Galactic plane are less massive on average
    }{
        The natal kicks applied to more massive BHs are suppressed as a result of fallback (Figure~\ref{fig:kicks-vs-mass}). Therefore, BHs found close to the Galactic plane, which have experienced weaker velocity kicks, are more massive on average (see Section~\ref{sec:correlations}).
    }

    \conclusion{%
        BH-star binaries have a bimodal BH mass distribution that is tightly correlated with orbital period and galactic location
    }{
        BH-star binaries can be split into two subclasses: (a) wide binaries that have not experienced Roche-lobe overflow, with high-mass BHs (since wide binaries can only survive if natal kicks are weak, disfavouring low-mass BHs) and (b) tight post-common-envelope binaries that can survive strong natal kicks imparted to low-mass BHs. The latter population have a significantly larger scale height as a result of their kicks (see Section~\ref{sec:bh_in_binaries}).
    }

    \conclusion{%
        The mass distribution of BHs in the Milky Way may provide a useful constraint on updated remnant mass prescriptions
    }{
        The Milky Way BH mass distribution is highly sensitive to remnant mass prescriptions (see Section~\ref{sec:mass-from-rmp}). Prescriptions from recent work considering the islands of explodability produce strong predictions of a dearth of BHs between 10 and 18 \msun, which may be testable by \romanShort or through X-ray observations (see Section~\ref{sec:discussion}).
    }

    \conclusion{%
        Accounting for the time-evolution of the Galactic potential more than doubles the escape fraction of BHs
    }{
        As a result of the mass growth of the Galaxy, BHs that formed earlier experienced a weaker gravitational potential. Accounting for this time evolution increases the escape fraction of BHs by a factor of $2.5\times$, while the scale height of BHs that remain bound to the Milky Way increases by around 20\% (see Section~\ref{sec:time-evolving-potential})
    }

    \conclusion{%
        Milky Way BH kinematics can constrain natal kick models
    }{
        BH kinematics (specifically their scale height and the correlation between BH mass and distance from the Galactic plane) are strongly impacted by assumptions regarding BH natal kicks (see Section~\ref{sec:kick-variations}). Observations may help to show how BH kicks differ from those applied to neutron stars.
    }

    \conclusion{%
        Neglecting binary interactions overestimates the scale height of BHs
    }{
        Evolving the fiducial population as single stars increases BH scale heights by 30\%. The scale height is smaller when accounting for binary evolution due to massive companions remaining bound to BHs after core-collapse events. The momentum imparted by the natal kick is shared by both members of the binary, leading to a lower net velocity for the system (see Section~\ref{sec:binary-vs-single}).
    }

    \conclusion{%
        The scale height and number of Milky Way BHs is robust to uncertainties in binary mass transfer
    }{
        Large variations in common-envelope efficiency, mass transfer efficiency, and mass transfer stability have negligible effects on our results (see Section~\ref{sec:mt-variations}).
    }
\end{conclusions}

Overall, the population of Milky Way BHs provides a unique window into binary interactions, supernova physics, and BH formation. With the upcoming \romanFull microlensing survey, results from ongoing and future X-ray surveys, and the release of \gaia DR4 we will soon have an exciting, rich dataset of Milky Way BHs that can be used to test and constrain our predictions.

\begin{acknowledgements}
    We thank David Sweeney for useful discussions related to this work, and for making the data from \citetalias{underworld} publicly available. We also thank Eric Bellm for helpful comments on an earlier draft of this work. Matt Orr provided helpful advice related to FIRE simulations that aided in our creation of the time-evolving potential model based on \firegal.
    The Flatiron Institute is funded by the Simons Foundation. T.W. and K.B. acknowledge support from NASA ATP grant 80NSSC24K0768.
    N.S.A. acknowledges support from the Heising-Simons Foundation under grant No.\ 2022-3542 and the H2H8 foundation.
\end{acknowledgements}

\software{This work made use of the following software packages: \texttt{astropy} \citep{astropy:2013,astropy:2018,astropy:2022}, \texttt{Jupyter} \citep{2007CSE.....9c..21P,kluyver2016jupyter}, \texttt{matplotlib} \citep{Hunter:2007}, \texttt{numpy} \citep{numpy}, \texttt{pandas} \citep{mckinney-proc-scipy-2010,pandas_18328522}, \texttt{python} \citep{python}, \texttt{scipy} \citep{2020SciPy-NMeth,scipy_18209846}, \texttt{COSMIC} \citep{COSMIC}, \texttt{Cython} \citep{cython:2011}, \texttt{Gala} \citep{Gala}, \texttt{h5py} \citep{collette_python_hdf5_2014,h5py_7560547}, and \texttt{tqdm} \citep{tqdm_18473238}.
This research has made use of the Astrophysics Data System, funded by NASA under Cooperative Agreement 80NSSC21M00561.
This research made use of \texttt{cogsworth} and its dependencies \citep{cogsworth_apjs, Wagg+2025:2025JOSS...10.7400W}.
Software citation information aggregated using \texttt{\href{https://www.tomwagg.com/software-citation-station/}{The Software Citation Station}} \citep{software-citation-station-paper,software-citation-station-zenodo}.}

\bibliographystyle{aasjournal}
\bibliography{bibs/paper, bibs/software}{}

\restartappendixnumbering

\allowdisplaybreaks

\appendix{}

\section{A simple time-evolving potential model for the Milky Way}\label{app:tep-model}

The exact mass (and morphology) evolution of the Milky Way remains highly uncertain \citep[e.g.,][]{Bland-Hawthorn+2016:2016ARA&A..54..529B}. Therefore, we use the \texttt{FIRE} simulations to emulate the evolution of the Milky Way \citep{Hopkins+2018:2018MNRAS.480..800H}. Specifically, we analyse \firegal, which most closely resembles the Milky Way within the currently available public simulations, and has been used to investigate Milky Way BH-BH binaries \citep[e.g.][]{Wetzel+2016, Lamberts+2018:2018MNRAS.480.2704L}. We track the mass growth of the inner $50 \unit{kpc}$ of the primary halo in \firegal over its past $12 \unit{Gyr}$ of evolution.

\begin{figure}
    \centering
    \includegraphics[width=\columnwidth]{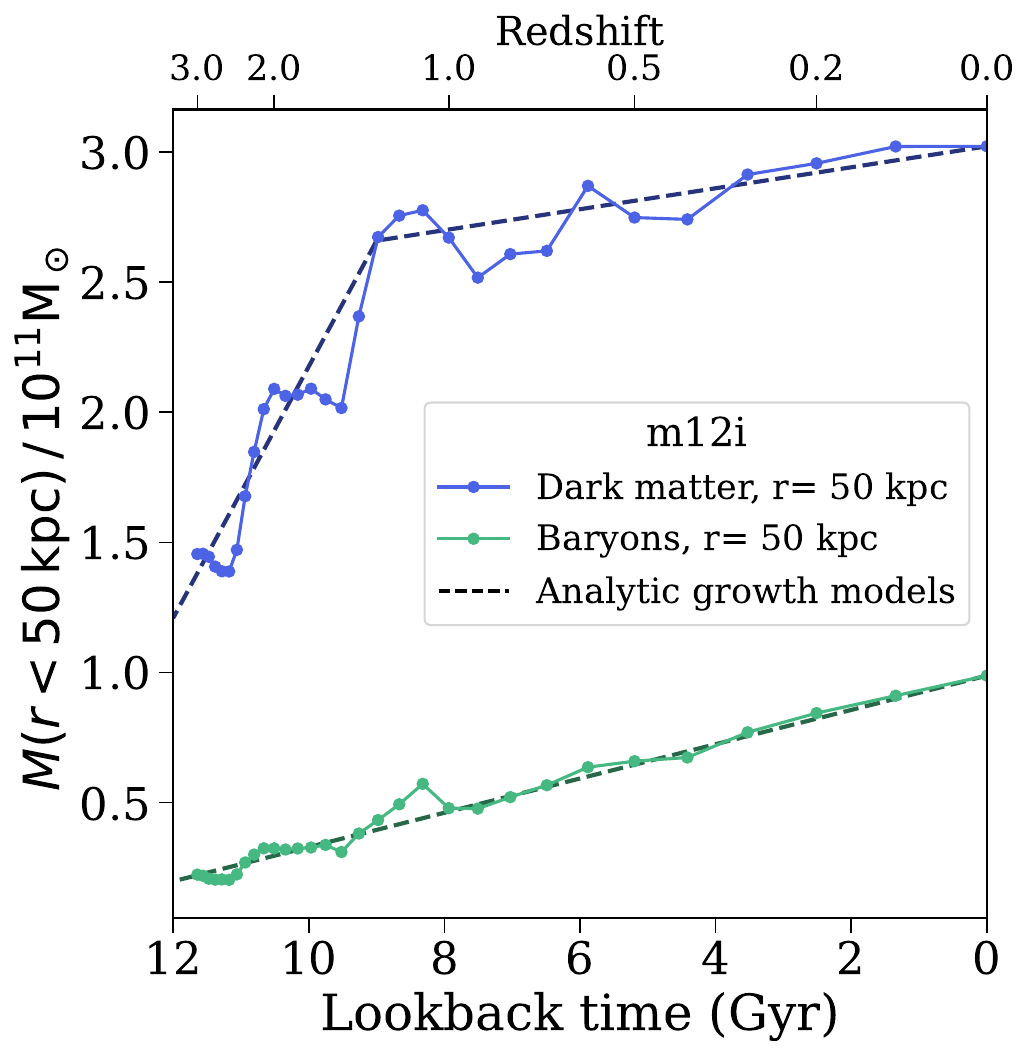}
    \caption{The time evolution of enclosed mass for \firegal can be approximated by analytic linear models. The coloured curves show the mass growth of dark matter (blue) and baryons (green) in the inner 50 kpc of the main halo of \firegal over the past $12 \, {\rm Gyr}$ (with the top axis showing the equivalent redshift). Dashed lines show our analytic models for the mass growth of each component. Baryons are well fit by a single linear model, dark matter is better approximated by two linear models, with a break at a lookback time of $9 \, {\rm Gyr}$.}
    \label{fig:m12i-mass-growth}
\end{figure}

We find that \firegal has significantly grown in mass, by more than a factor of two, with most of the growth occurring in the first $3 \unit{Gyr}$. In Figure~\ref{fig:m12i-mass-growth}, we show the evolution of the mass enclosed within the central $50\unit{kpc}$ of \firegal's main halo. We calculate the enclosed mass at snapshots between redshift $z=0$ and $z=3$ in steps of $0.1$. The non-monotonic evolution of the mass is because satellites pass through the inner $50 \unit{kpc}$ of the galaxy, resulting in the occasional apparent fluctuations in mass.

The dark matter mass growth follows the typical `two-phase' evolution, which is characterised by a rapid early phase (seen here in the first 3 Gyr of \firegal's evolution) and a later extended phase \citep{Oser+2010:2010ApJ...725.2312O}. In contrast, the baryonic mass follows a simple linear growth.

We approximate the mass growth of the main halo of \firegal with simple analytic, linear models (shown as dashed lines in Figure~\ref{fig:m12i-mass-growth}). For the dark matter growth we use a two-part linear model, such that the mass of dark matter, $M_{\rm DM}$, is given by
\begin{equation}\label{eq:dm_mass_growth}
    M_{\rm DM} = \begin{cases}
        M_{\rm DM, f} (2.32 - \frac{0.16 \tau}{\unit{Gyr}}) & \tau > 9 \unit{Gyr} \\
        M_{\rm DM, f} (1 - \frac{0.12 \tau}{9 \unit{Gyr}}) & \tau \le 9 \unit{Gyr} \\
    \end{cases},
\end{equation}
where $M_{\rm DM, f}$ is the final dark matter mass and $\tau$ is the lookback time. For the baryonic mass, $M_{\rm b}$, we use a single linear model:
\begin{equation}\label{eq:baryon_mass_growth}
    M_{\rm b} = M_{\rm b, f} \qty(1 - \frac{0.8 \tau}{12 \unit{Gyr}}),
\end{equation}
where $M_{\rm b, f}$ is the final baryonic mass.

We use this model to construct a gravitational potential for the Milky Way that increases in mass over time. We assume that the present-day state of this potential is identical to the (static) fiducial model, which is the \texttt{MilkyWayPotential} from \gala. However, in the time-evolving potential, we alter the dark matter halo component, which follows an NFW profile \citep{Navarro+1997:1997ApJ...490..493N}, such that its mass follows Eq.~\ref{eq:dm_mass_growth}, with $M_{\rm DM, f}$ equal to the dark matter mass we assume in the static case. Similarly, we alter the disc component, which is a sum of three Miyamoto-Nagai disc potentials \citep{Miyamoto+1975:1975PASJ...27..533M, Smith+2015:2015MNRAS.448.2934S}, such that its mass follows Eq.~\ref{eq:baryon_mass_growth}, with $M_{\rm b, f}$ equal to the disc mass we assume in the static case.

\section{Analytic estimate of impact of a time-evolving potential on BH scale heights}\label{app:analytic-tep}

One can estimate the effect of a time-evolving potential on the vertical kinematics of stars through an analytic approximation. Consider two stars initially at the midplane $z=0$ with zero initial vertical velocity that each receive a vertical kick $\delta_{v_z}$ that is much larger than the velocity dispersion (i.e.\ $\delta_{v_z} \gg \sigma_{v_z}$). The first star, A, receives this kick early in the history of the galaxy at $t = - 12 \unit{Gyr}$, while the other, B, is kicked at present day. Let us assume that the disc surface mass density, $\Sigma$, has grown by a factor $f$ in the past 12 Gyr.

The vertical frequency, $\Omega_z$, sets the maximum vertical distance, $z_{\rm max}$, such that
\begin{equation}
    z_{\rm max} \propto \frac{\delta_{v_z}}{\Omega_z}.
\end{equation}
Since we are assuming a factor of $f$ growth in $\Sigma$ (and $\Omega_z \propto \sqrt{\Sigma}$) the frequency in the past was
\begin{equation}
    \Omega_z(t = -12\unit{Gyr}) \propto \frac{\Omega_{z,0}}{\sqrt{f}},
\end{equation}
where $\Omega_{z,0}$ is the present vertical frequency.

Shortly after receiving its kick, star A will reach a height of $z_{\rm max} \propto \sqrt{f} \, \delta_{v_z} / \Omega_{z,0}$. However, the potential deepens adiabatically over the intervening time. We know that the vertical action is invariant to adiabatic changes and thus
\begin{equation}
    J_z \sim \Omega_z \, z_{\rm max}^2
\end{equation}
does not change. So we can state that
\begin{equation}
    J_{z, {\rm past, A}} \sim \frac{\Omega_{z,0}}{\sqrt{f}} \cdot \frac{f \, v_z^2}{\Omega_{z,0}^2} = \frac{\sqrt{f} \, v_z^2}{\Omega_{z, 0}},
\end{equation}
and
\begin{equation}
    J_{z, {\rm A}} \sim \Omega_{z,0} \, z_{\rm max, A}^2,
\end{equation}
such that solving for the maximum height at present day we now see
\begin{equation}
    z_{\rm max, A} = f^{1/4} \, \frac{\delta_{v_z}}{\Omega_0}.
\end{equation}
For star B, its maximum height is simply
\begin{equation}
    z_{\rm max, B} = \frac{\delta_{v_z}}{\Omega_{z,0}},
\end{equation}
and thus we find that
\begin{equation}
    z_{\rm max, A} / z_{\rm max, B} = f^{1/4}.
\end{equation}

For our model of the mass growth of the galaxy, the surface mass density has increased by a factor of $f = 2.5$ since the birth of the Milky Way. Therefore a star born at the start of the Milky Way would have a scale height that is a factor of $1.26$ ($=2.5^{1/4}$) larger than a star that received the same kick around present day. Of course not all stars are formed at the start of the Milky Way, so in reality we'd expect the factor to be somewhere between $1$ and $1.26$. This is in agreement with our finding of a 20\% increase in the \texttt{cogsworth} simulation, which reflects the fact that star formation is skewed to earlier times.

\section{Comparison to previous work}\label{sec:compare}

Earlier works have investigated the Milky Way population of BHs. Most recently, \citetalias{Olejak+2020:2020A&A...638A..94O} presented a detailed analysis of the demographics of BHs, without integrating orbits through a galactic potential. \citetalias{underworld} instead focused on the spatial distribution of BHs and NSs, integrating orbits through a Milky Way potential, but did not perform population synthesis. We compare our results to these works in turn below.

\paragraph{\citet{underworld}}

\citetalias{underworld} presented predictions for the ``Galactic Underworld'', estimating spatial distribution of compact objects in the Milky Way, using \texttt{GALAXIA} \citep{Sharma+2011:2011ApJ...730....3S} simulations, integrated through the galaxy using \texttt{galpy} \citep{Bovy+2015:2015ApJS..216...29B}.

We broadly agree with \citetalias{underworld} qualitatively; BHs exhibit a significantly different spatial structure than the visible galaxy, but only a small fraction escape the galaxy. They report that BHs have a scale height of $900\pm40\unit{pc}$. However, this scale height is calculated assuming an exponential profile for the BHs and was only calculated in the solar neighbourhood, specifically $7.5 < R / \unit{kpc} < 8.5$ where $R$ is Galactocentric radius (D.\ Sweeney, priv.\ comm.). We find that an exponential profile is a poor fit for the distribution of BHs, which have a much longer tail in $z$. We used the publicly available data from \citetalias{underworld} to recalculate the scale height of their simulated BHs. When using an exponential profile to calculate the scale height for the full disc (using the larger sample of ${\sim}74,000$ BHs rather than the ${\sim}2300$ found in the solar neighbourhood) we find a scale height of $590\unit{pc}$. However, when applying our method that doesn't assert an exponential profile, we find a scale height of $850\unit{pc}$. This scale height is more comparable to our reported value of $790\unit{pc}$ (calculated from $2\times10^6$ simulated BHs). Therefore, our results are broadly consistent with \citetalias{underworld}, but we find that the scale height of BHs is slightly lower than what they report, and that the distribution of BHs is not well described by an exponential profile.

\paragraph{\citet{Olejak+2020:2020A&A...638A..94O}}

\citetalias{Olejak+2020:2020A&A...638A..94O} performed a detailed population synthesis study of the Milky Way BH population, using the \texttt{StarTrack} code \citep{Belczynski+2008}. Their study focused on the demographics of the BH population, and as such they did not consider the impact of the potential on the kinematics of BHs. They use a simpler star formation history for the Galaxy, assuming mostly constant star formation in different components, with increases in metallicity over time.

They find that the total number of BHs in the Milky Way is around $1.29 \times 10^8$, which is around 25\% lower than our fiducial prediction of $1.73 \times 10^8$. However, they assume the \citet{Fryer+2012:2012ApJ...749...91F} \textit{rapid} remnant mass prescription, which we find produces around $1.34 \times 10^8$ BHs, strikingly close to their prediction. The small difference between our prediction for the \textit{rapid} prescription and their prediction is likely due to differences in the star formation history and metallicity evolution of the Galaxy that we assume, as well as differences in the population synthesis code used.

Our predictions for the numbers of different types of BH binaries are broadly consistent with \citetalias{Olejak+2020:2020A&A...638A..94O}. They predict around $9.3 \times 10^6$ total BHs in binaries, with $7.3 \times 10^6$ BH-BHs, $2.7 \times 10^5$ BH-NSs, $1.7 \times 10^6$ BH-WDs, and $1.7 \times 10^5$ BH-star binaries. Our most similar variation to their model (the \textit{rapid} remnant mass prescription) predicts similar, though slightly higher numbers: \num{8.6e6} BH-BHs, \num{3.5e5} BH-NSs, \num{2.5e6} BH-WDs, and \num{3.8e5} BH-star binaries. The exception is BH-star binaries which we find at twice the rate of \citetalias{Olejak+2020:2020A&A...638A..94O}.

\section{BH velocity distributions}\label{app:velocities}

\begin{figure}
    \centering
    \includegraphics[width=\columnwidth]{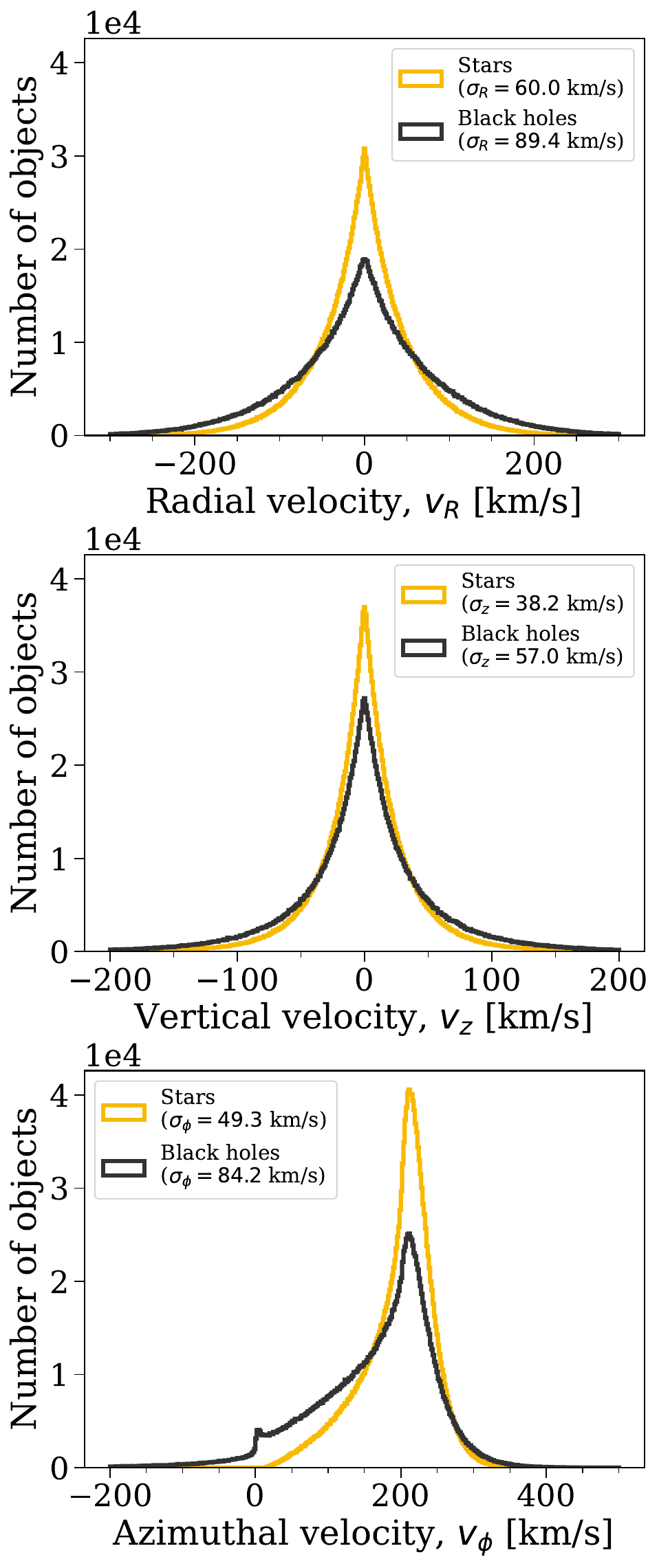}
    \caption{The typical velocity dispersion for BHs is 50-70\% larger than stars in our fiducial model. The stellar distributions are drawn from the star formation history model. Each panel shows the distribution of one component velocity.}
    \label{fig:vel_components}
\end{figure}

In Figure~\ref{fig:vel_components} we show the component velocities for Milky Way BHs in the fiducial model, compared to the stellar population (as drawn from the star formation history). As noted in the legend, the velocity dispersion increases by 50\% for the radial and vertical velocity, and increases by 70\% for the azimuthal velocity. This additional increase is driven by the tail to negative azimuthal velocities. These retrograde BHs (those with negative $v_{\phi}$) represent approximately 4\% of the BHs that remain bound to the Milky Way.

\begin{figure*}
    \centering
    \includegraphics[width=\textwidth]{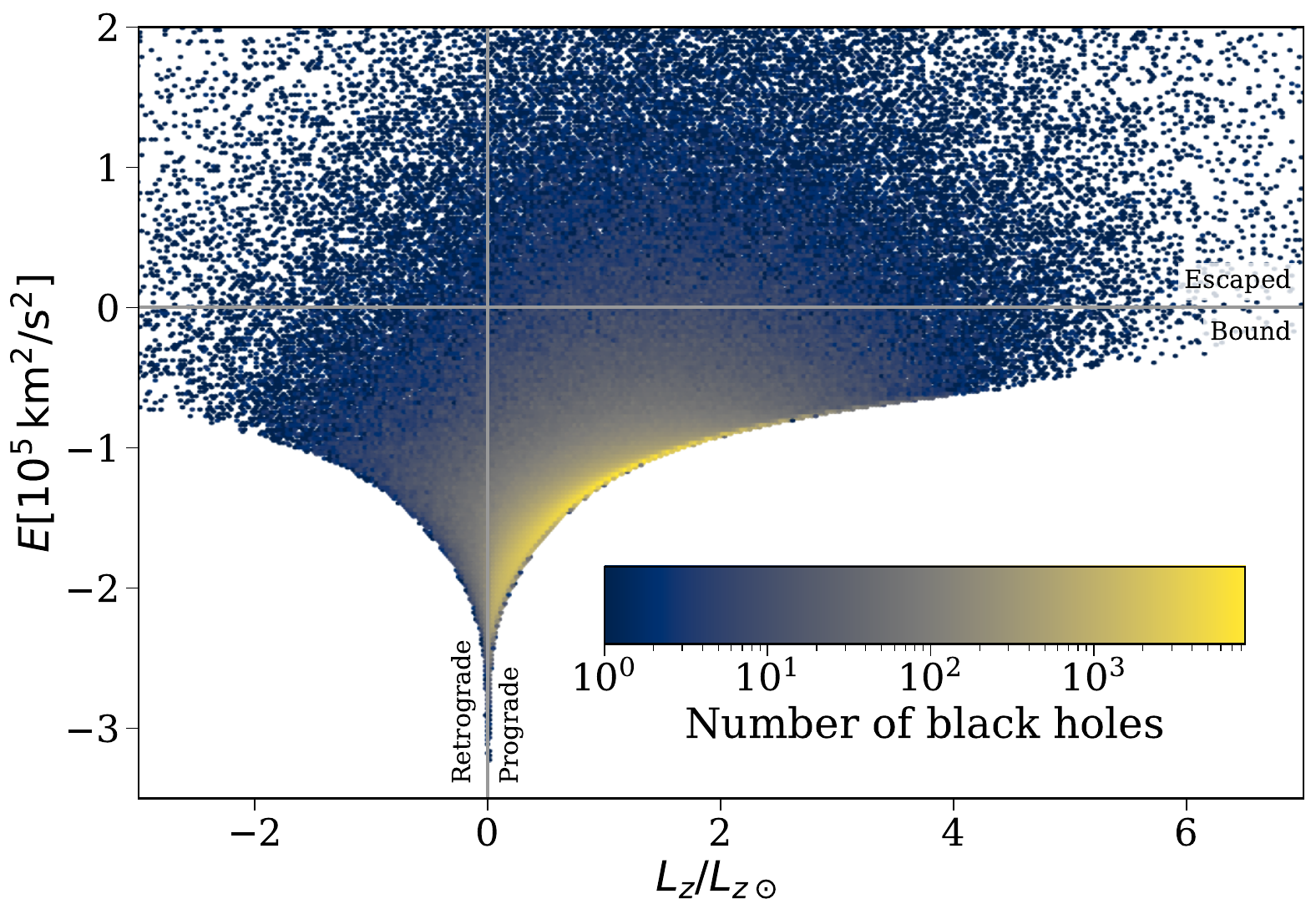}
    \caption{Most BHs have Galactic orbits that are disc-like, but broader than the stellar population. However, small subsets of BHs have retrograde orbits or have escaped the galaxy. We show the vertical angular momentum (scaled to solar values), $L_z$, against the Galactic orbital energy. Grey lines separate prograde and retrograde orbits (positive and negative $L_z$) and orbits that are bound to the Galaxy or escaped (negative and positive $E$).}
    \label{fig:vel_Lz_E}
\end{figure*}

We explore the Galactic orbits of BHs in more detail in Figure~\ref{fig:vel_Lz_E}, which shows the vertical angular momentum ($L_z$) vs. ($E$). From this figure, we can see that the majority of BHs follow a narrow contour in $L_z-E$, corresponding to prograde, circular orbits. The distribution is broader than one would expect for a typical stellar distribution, which is driven by the natal kicks experienced by the BHs. The BH orbits also extend into more radial, halo-like orbits, unlike the stars in the disc (where the BHs were formed in our model). As we have noted in earlier sections, a fraction of BHs escape the galaxy (this population extends much further than the limits of this figure). Moreover, there are a subset of BHs with retrograde orbits (negative $L_z$) around the Milky Way. Although many of these are more radial orbits, there are some BHs which are following retrograde, \textit{disc-like} orbits due to lucky (and strong) natal kicks.

\section{Robustness to variations in binary mass transfer physics}\label{app:mt_variations}

\begin{table*}
    \centering
    \begin{tabular}{lcccccccc}
\toprule
Model & $N_{\rm BH} / 10^8$ & $M_{\rm BH} / 10^9 M_\odot$ & $h_{z, {\rm eff}}$ (pc) & $f_{\rm esc}$ & $\median{m_{|z| < 1 \, {\rm kpc}}}$ & $\median{m_{|z| \geq 1 \, {\rm kpc}}}$ & $f_{\rm bound}$ & $f_{\rm merger}$ \\
\midrule
\textbf{Fiducial} & $1.73$ & $1.50$ & $786$ & 2.7\% & $7.70$ & $6.22$ & $9\%$ & $33\%$ \\
\addlinespace[0.5em]
\multicolumn{8}{l}{\textbf{Mass transfer efficiency}} \\
\quad $\beta = 0.0$ & $1.37$ & $1.21$ & $771$ & 2.6\% & $7.91$ & $6.05$ & $11\%$ & $38\%$ \\
\quad $\beta = 0.5$ & $1.53$ & $1.34$ & $790$ & 2.7\% & $7.83$ & $6.10$ & $10\%$ & $35\%$ \\
\quad $\beta = 1.0$ & $1.73$ & $1.50$ & $786$ & 2.7\% & $7.69$ & $6.21$ & $9\%$ & $33\%$ \\
\addlinespace[0.5em]
\multicolumn{8}{l}{\textbf{Common-envelope efficiency}} \\
\quad $\alpha_{\rm CE} = 0.1$ & $1.77$ & $1.53$ & $775$ & 2.7\% & $7.65$ & $6.26$ & $8\%$ & $35\%$ \\
\quad $\alpha_{\rm CE} = 0.5$ & $1.75$ & $1.52$ & $779$ & 2.7\% & $7.67$ & $6.24$ & $9\%$ & $34\%$ \\
\quad $\alpha_{\rm CE} = 2.0$ & $1.71$ & $1.48$ & $799$ & 2.7\% & $7.72$ & $6.12$ & $9\%$ & $32\%$ \\
\quad $\alpha_{\rm CE} = 10.0$ & $1.65$ & $1.42$ & $800$ & 2.7\% & $7.77$ & $6.02$ & $10\%$ & $28\%$ \\
\addlinespace[0.5em]
\multicolumn{8}{l}{\textbf{Mass transfer stability}} \\
\quad $q_{\rm crit, B} = 0$ & $1.91$ & $1.74$ & $823$ & 2.9\% & $8.42$ & $6.32$ & $4\%$ & $65\%$ \\
\quad $q_{\rm crit, B} = \infty$ & $1.78$ & $1.55$ & $815$ & 2.8\% & $7.74$ & $6.09$ & $13\%$ & $27\%$ \\
\bottomrule
\end{tabular}
    \caption{As Table~\ref{tab:summary_table}, but for mass transfer physics variations.}
    \label{tab:mt_summary_table}
\end{table*}

We explored the sensitivity of our results to uncertainties in binary mass transfer. Specifically, we varied (1) the stability of mass transfer in two bounding variations (in one all mass transfer is stable, in another, all mass transfer is unstable), (2) the efficiency of stable mass transfer (we varied this to 0, 0.5, and 1.0), and (3) the efficiency of common-envelope phases (we varied this to 0.1, 0.5, and 1.0). We find that these variations have a negligible effect on the scale height of BHs and only a small effect on the number of BHs produced (see Table~\ref{tab:mt_summary_table}). The main deviation from the fiducial model comes from low stable mass transfer efficiency, which produces a lower number of BHs because secondaries cannot accrete enough to form BHs. Moreover, the correlation between mean BH mass and Galactic height remains present in all variations (see Figure~\ref{fig:mean_bh_mass_vs_z_MT_variations}).

\begin{figure}
    \centering
    \includegraphics[width=\columnwidth]{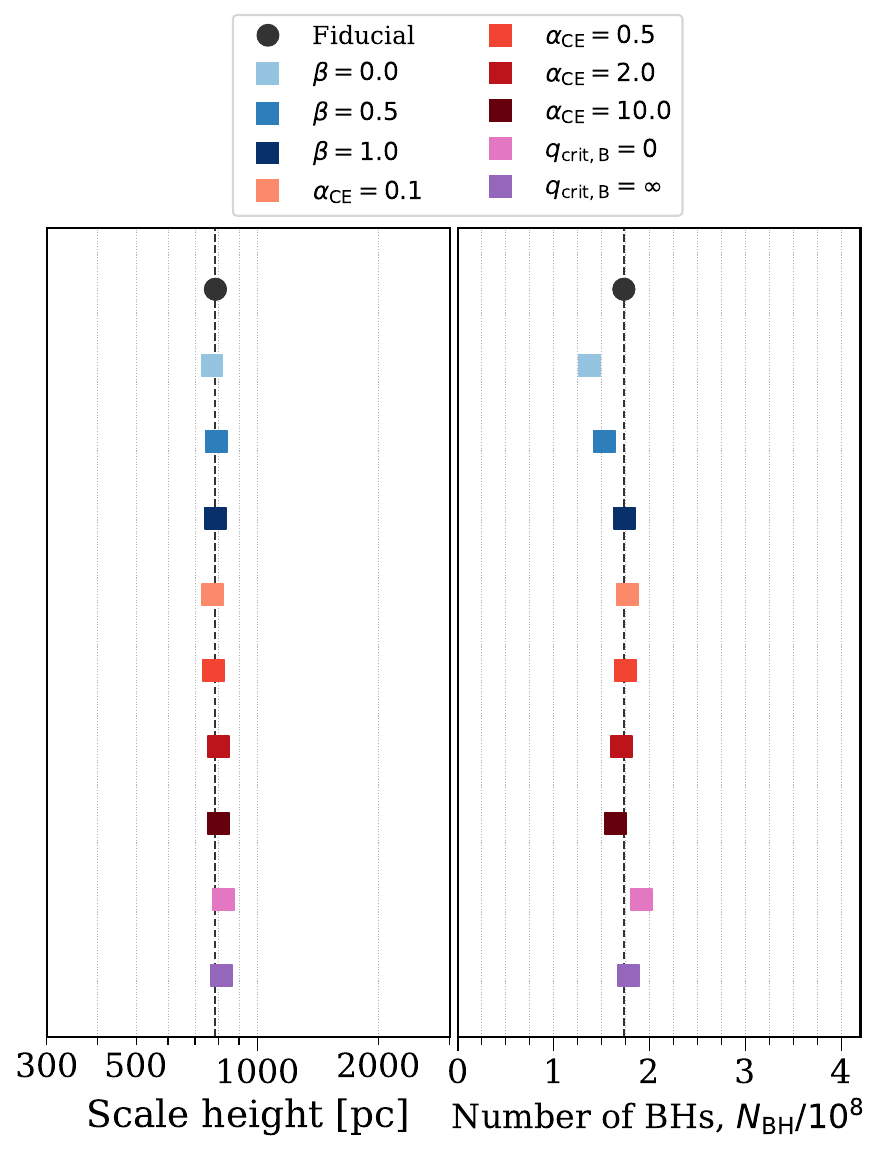}
    \caption{The scale height and number of BHs produced is robust to uncertainties in mass transfer physics. As Figure~\ref{fig:summary-stats}, but showing mass transfer variations (with the same axes limits).}
    \label{fig:summary-stats-mt}
\end{figure}

\begin{figure}
    \centering
    \includegraphics[width=\columnwidth]{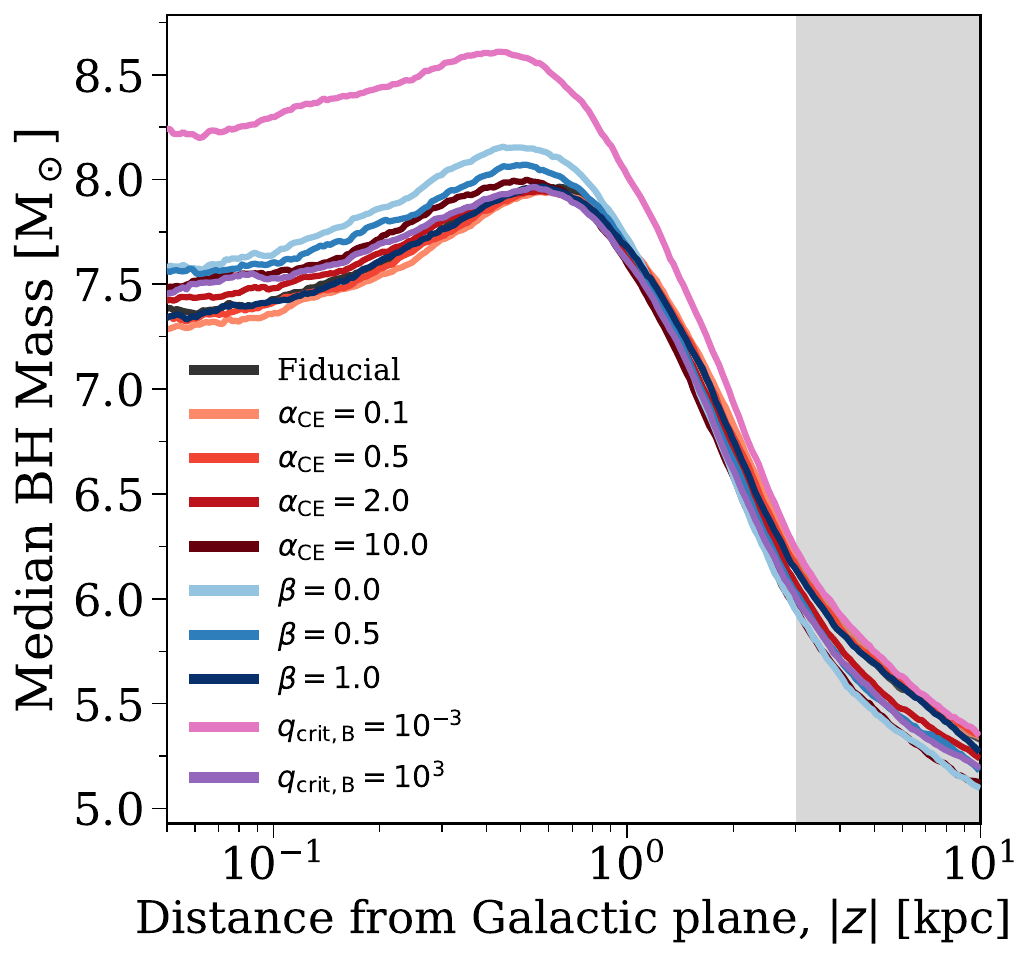}
    \caption{Correlations between mean BH mass and Galactic height are robust to variations in binary mass transfer physics. As Figure~\ref{fig:average-mass-by-z}, but showing lines for binary physics variations (described in Appendix~\ref{app:mt_variations}). Although the absolute scale changes (particularly for the variation in which mass transfer is forced to always be unstable), the trend of an increase in Galactic height leading to a decrease in average BH mass remains.}
    \label{fig:mean_bh_mass_vs_z_MT_variations}
\end{figure}

\section{Supplementary tables and figures}

\begin{sidewaysfigure*}
    \centering
    \includegraphics[width=0.95\textheight]{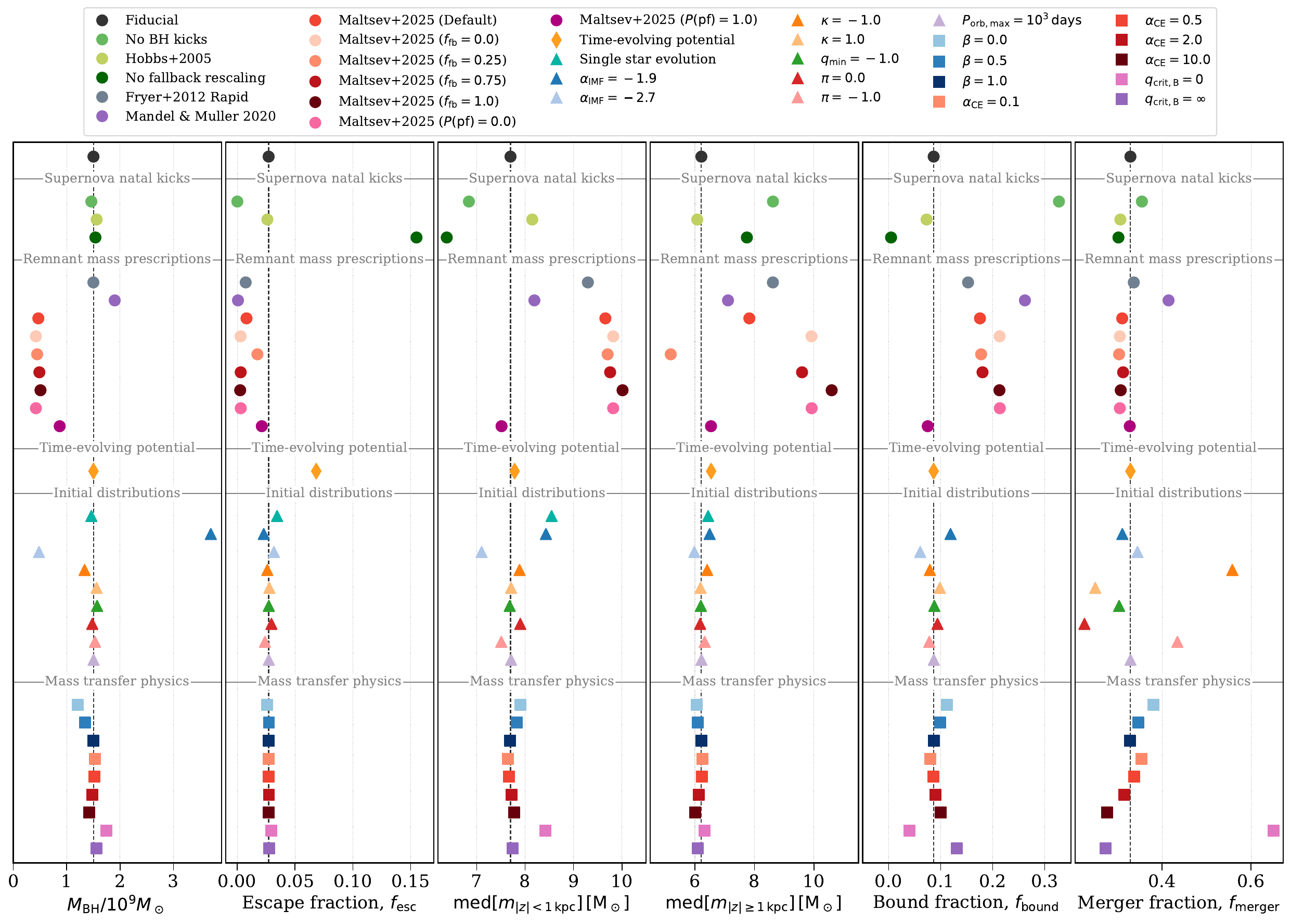}
    \caption{As Figure~\ref{fig:summary-stats}, but for the other columns reported in Table~\ref{tab:summary_table}.}
    \label{fig:summary-other-cols}
\end{sidewaysfigure*}

\begin{figure}
    \centering
    \includegraphics[width=\columnwidth]{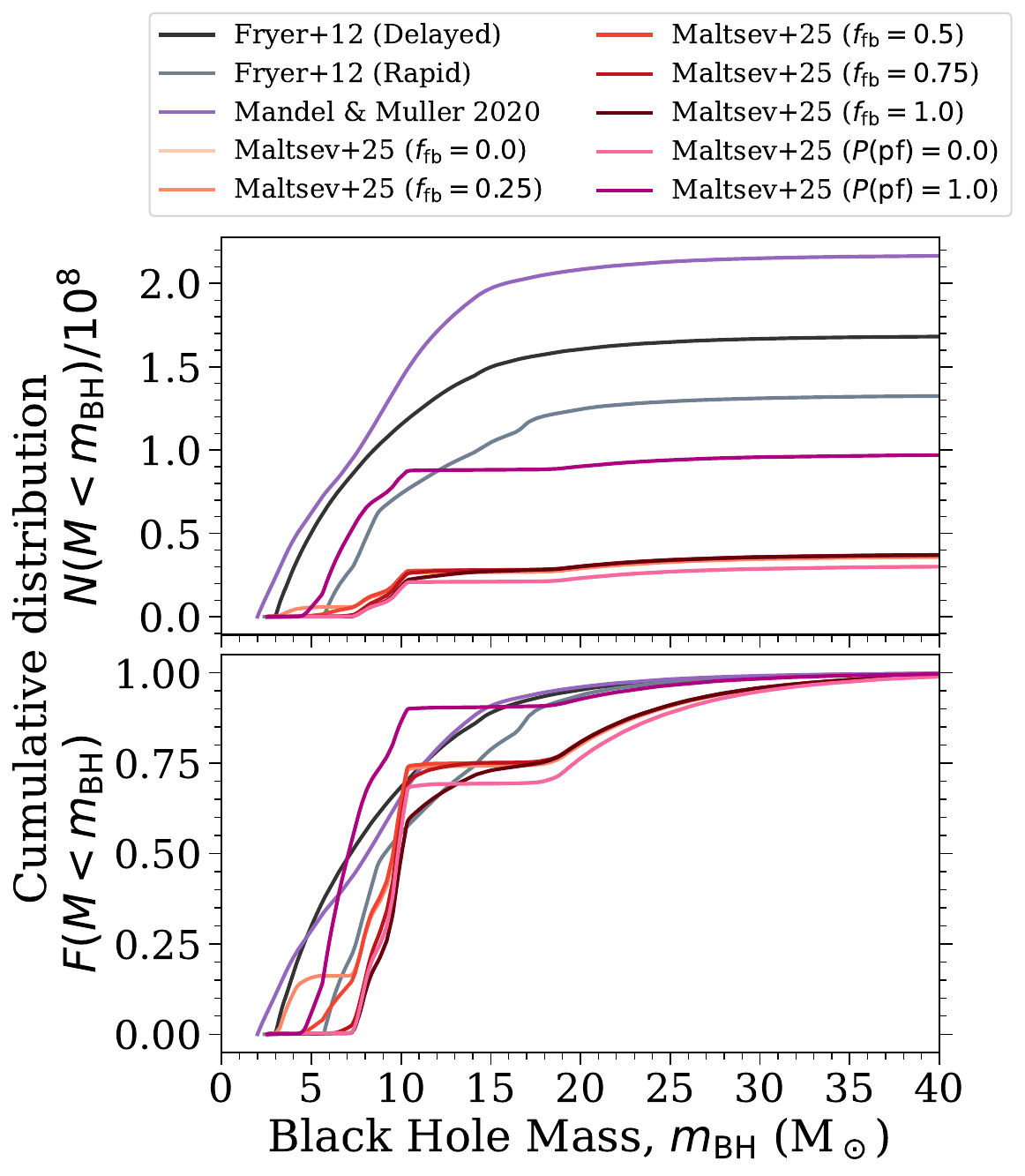}
    \caption{Cumulative distributions for the masses of Milky Way BHs for each variation of remnant mass prescription that we consider. The top panel shows the cumulative \textit{number} of BHs above a given mass, while the bottom panel shows the cumulative \textit{fraction}.}
    \label{fig:rmp-cdfs}
\end{figure}

Figure~\ref{fig:summary-other-cols} vizualizes the information in Table~\ref{tab:summary_table}, in the same manner as Figure~\ref{fig:summary-stats}.
Table~\ref{tab:bh_binaries} shows the number and average mass of BHs in different types of binaries at present day in the Milky Way for each of the variations. The total numbers are presented visually in Figure~\ref{fig:bh-binary-rates}.
Figure~\ref{fig:rmp-cdfs} shows the cumulative distribution of the masses of BHs in the Milky Way (similar to the distributions in Figure~\ref{fig:bh_mass_rmp}).

\begin{table*}[htbp]
    \centering
    \begin{tabular}{lcccc}
\toprule
Model & $N_{\rm BH\text{-}star} / 10^4$ & $N_{\rm BH\text{-}WD} / 10^5$ & $N_{\rm BH\text{-}NS} / 10^5$ & $N_{\rm BH\text{-}BH} / 10^6$ \\
\midrule
\textbf{Fiducial} & $17.1$ & $12.7$ & $1.4$ & $6.7$ \\
\addlinespace[0.5em]
\multicolumn{4}{l}{\textbf{Supernova natal kicks}} \\
\quad No BH kicks & $74.3$ & $56.4$ & $4.0$ & $24.3$ \\
\quad Hobbs+2005 & $15.5$ & $10.8$ & $0.5$ & $5.9$ \\
\quad No fallback rescaling & $2.2$ & $2.8$ & $0.5$ & $0.2$ \\
\addlinespace[0.5em]
\multicolumn{4}{l}{\textbf{Remnant mass prescriptions}} \\
\quad Fryer+2012 Rapid & $38.0$ & $24.8$ & $3.5$ & $8.6$ \\
\quad Mandel\&Muller2020 & $102.8$ & $80.0$ & $3.2$ & $23.7$ \\
\quad Maltsev+2025 & $15.0$ & $10.6$ & $3.7$ & $2.5$ \\
\multicolumn{4}{l}{\textit{\quad Maltsev+2025 fallback mass fraction}} \\
\quad\quad $f_{\rm fb} = 0.0$ & $15.1$ & $10.5$ & $3.9$ & $2.5$ \\
\quad\quad $f_{\rm fb} = 0.25$ & $15.1$ & $10.6$ & $3.6$ & $2.5$ \\
\quad\quad $f_{\rm fb} = 0.75$ & $15.2$ & $10.8$ & $4.1$ & $2.6$ \\
\quad\quad $f_{\rm fb} = 1.0$ & $19.1$ & $13.0$ & $4.6$ & $3.0$ \\
\multicolumn{4}{l}{\textit{\quad Maltsev+2025 partial fallback probability}} \\
\quad\quad $P({\rm pf}) = 0.0$ & $15.0$ & $10.5$ & $4.0$ & $2.5$ \\
\quad\quad $P({\rm pf}) = 1.0$ & $15.7$ & $11.6$ & $2.5$ & $3.0$ \\
\addlinespace[0.5em]
\multicolumn{4}{l}{\textbf{Galactic gravitational potential}} \\
\quad Time-evolving potential & $17.1$ & $12.7$ & $1.4$ & $6.7$ \\
\addlinespace[0.5em]
\multicolumn{4}{l}{\textbf{Initial mass function}} \\
\quad $\alpha_{\rm IMF} = -1.9$ & $47.2$ & $34.5$ & $3.7$ & $21.2$ \\
\quad $\alpha_{\rm IMF} = -2.7$ & $4.1$ & $3.5$ & $0.4$ & $1.6$ \\
\addlinespace[0.5em]
\multicolumn{4}{l}{\textbf{Initial mass ratio distribution}} \\
\quad $\kappa = -1.0$ & $460.6$ & $38.2$ & $0.9$ & $1.8$ \\
\quad $\kappa = 1.0$ & $0.7$ & $2.5$ & $0.9$ & $8.7$ \\
\quad $q_{\rm min} = -1.0$ & $0.8$ & $5.4$ & $1.5$ & $7.6$ \\
\addlinespace[0.5em]
\multicolumn{4}{l}{\textbf{Initial orbital period distribution}} \\
\quad $\pi = 0.0$ & $21.4$ & $16.4$ & $1.6$ & $7.0$ \\
\quad $\pi = -1.0$ & $12.9$ & $9.7$ & $1.2$ & $6.3$ \\
\quad $P_{\rm orb, max} = 10^3\,{\rm days}$ & $17.1$ & $12.5$ & $1.4$ & $6.7$ \\
\addlinespace[0.5em]
\multicolumn{4}{l}{\textbf{Mass transfer efficiency}} \\
\quad $\beta = 0.0$ & $16.1$ & $13.3$ & $5.5$ & $6.7$ \\
\quad $\beta = 0.5$ & $16.2$ & $12.3$ & $1.8$ & $6.8$ \\
\quad $\beta = 1.0$ & $16.7$ & $13.1$ & $1.4$ & $6.7$ \\
\addlinespace[0.5em]
\multicolumn{4}{l}{\textbf{Common-envelope efficiency}} \\
\quad $\alpha_{\rm CE} = 0.1$ & $14.8$ & $10.5$ & $1.1$ & $6.4$ \\
\quad $\alpha_{\rm CE} = 0.5$ & $16.5$ & $12.7$ & $1.4$ & $6.7$ \\
\quad $\alpha_{\rm CE} = 2.0$ & $17.7$ & $14.5$ & $1.6$ & $6.8$ \\
\quad $\alpha_{\rm CE} = 10.0$ & $21.5$ & $17.6$ & $2.1$ & $7.1$ \\
\addlinespace[0.5em]
\multicolumn{4}{l}{\textbf{Mass transfer stability}} \\
\quad $q_{\rm crit, B} = 0$ & $17.1$ & $9.2$ & $0.5$ & $3.2$ \\
\quad $q_{\rm crit, B} = \infty$ & $21.7$ & $27.4$ & $8.6$ & $9.8$ \\
\bottomrule
\end{tabular}
    \caption{Similar to Table~\ref{tab:summary_table}, but showing statistics of BH binaries at present day in the Milky Way for each of the variations. The columns show the number and average BH mass of BH-star, BH-WD, BH-NS, and BH-BH binaries.}
    \label{tab:bh_binaries}
\end{table*}

\end{document}